\documentclass[journal abbreviation, manuscript]{copernicus}

\usepackage{tabularx}
\usepackage{multirow}
\usepackage{supertabular}
\usepackage{booktabs}
\usepackage{amsmath}

\usepackage{caption}
\usepackage{subcaption}
\usepackage{longtable}
\usepackage{pdflscape}

\begin{document}

\title{Discovery of Spherules of Likely Extrasolar Composition in the Pacific Ocean Site of the CNEOS 2014-01-08 (IM1) Bolide}

\Author[1,2]{Abraham}{Loeb}
\Author[2]{Toby}{Adamson}
\Author[2]{Sophie}{Bergstrom}
\Author[1,2]{Richard}{Cloete}
\Author[2,7]{Shai}{Cohen}
\Author[2]{Kevin}{Conrad}
\Author[1,2]{Laura}{Domine}
\Author[2,3]{Hairuo}{Fu}
\Author[2]{Charles}{Hoskinson}
\Author[2,3]{Eugenia}{Hyung}
\Author[2,3]{Stein}{Jacobsen}
\Author[2]{Mike}{Kelly}
\Author[2]{Jason}{Kohn}
\Author[2]{Edwin}{Lard}
\Author[2,4]{Sebastian}{Lam}
\Author[2,6]{Frank}{Laukien}
\Author[2,5]{Jim}{Lem}
\Author[2]{Rob}{McCallum}
\Author[2]{Rob}{Millsap}
\Author[2,3]{Christopher}{Parendo}
\Author[2,3]{Michail}{Pataev}
\Author[2,4]{Chaitanya}{Peddeti}
\Author[2]{Jeff}{Pugh}
\Author[2,7]{Shmuel}{Samuha}
\Author[1,2]{Dimitar}{Sasselov}
\Author[2]{Max}{Schlereth}
\Author[2]{J.J.}{Siler}
\Author[1,2]{Amir}{Siraj}
\Author[2]{Peter Mark}{Smith}
\Author[2]{Roald}{Tagle}
\Author[2]{Jonathan}{Taylor}
\Author[2,4]{Ryan}{Weed}
\Author[2]{Art}{Wright}
\Author[2]{Jeff}{Wynn}
\affil[1]{Department of Astronomy, Harvard University, 60 Garden Street, Cambridge, MA 02138, USA}
\affil[2]{Interstellar Expedition of the Galileo Project, 60 Garden Street, Cambridge, MA 02138, USA}
\affil[3]{Department of Earth and Planetary Sciences, Harvard University, 20 Oxford Street, Cambridge, MA 02138, USA}
\affil[4]{Department of Nuclear Engineering, UC Berkeley, 4153 Etcheverry Hall, MC 1730, Berkeley, CA 94720, USA}
\affil[5]{Department of Mining Engineering, PNG University of Technology, Lae 411, Papua New Guinea}
\affil[6]{Department of Chemistry and Chemical Biology, Harvard University, 12 Oxford Street, Cambridge, MA 02138, USA}
\affil[7]{Department of Materials Engineering, NRCN, P.O. Box 9001, Beer-Sheva 84190, Israel}




\correspondence{Abraham Loeb, Head of the Galileo Project (aloeb@cfa.harvard.edu)}

\runningtitle{Discovery of IM1 spherules}

\runningauthor{Interstellar Expedition team}

\received{}
\pubdiscuss{} 
\revised{}
\accepted{}
\published{}


\firstpage{1}

\maketitle

\nolinenumbers
\begin{abstract}
We have conducted an extensive towed-magnetic-sled survey during the period 14-28 June, 2023, over the seafloor about 85 km north of Manus Island, Papua New Guinea, and found about 700 spherules of diameter 0.05-1.3 millimeters in our samples, of which 57 were analyzed so far. Approximately $0.26~{\rm km^2}$ of seafloor was sampled in this survey, centered around the calculated path of the bolide CNEOS 2014-01-08 (IM1) with control areas north and south of that path. The spherules, significantly concentrated along the expected meteor path, were retrieved from seafloor depths ranging between 1.5-2.2 km. Mass spectrometry of 47 spherules near the high-yield regions along IM1's path reveals a distinct extra-solar abundance pattern for 5 of them, while background spherules have abundances consistent with a solar system origin. The unique spherules show an excess of Be, La and U, by up to three orders of magnitude relative to the solar system standard of CI chondrites. These ``BeLaU"-type
spherules, never seen before, also have very low refractory siderophile elements such as Re. Volatile elements, such as Mn, Zn, Pb, are depleted
as expected from evaporation losses during a meteor's airburst. In addition, the mass-dependent variations in $^{57}$Fe/$^{54}$Fe and $^{56}$Fe/$^{54}$Fe are also consistent with evaporative loss of the light isotopes during the spherules' travel in the atmosphere.  The ``BeLaU" abundance pattern is not found in control regions outside of IM1's path and does not match commonly manufactured alloys or natural meteorites in the solar system. This evidence points towards an association of ``BeLaU"-type spherules with IM1, supporting its interstellar origin independently of the high velocity and unusual material strength implied from the CNEOS data. We suggest that the ``BeLaU" abundance pattern could have originated from a highly differentiated magma ocean of a planet with an iron core outside the solar system or from more exotic sources.
\end{abstract}

\introduction  
On 8 January 2014 US government satellite sensors detected three atmospheric detonations in rapid succession about 84~km north of Manus Island, outside the territorial waters of Papua New Guinea (20 km)~\footnote{\href{[https://www.un.org/depts/los/LEGISLATIONANDTREATIES/PDFFILES/PNG_1977_Act7.pdf]}{https://www.un.org/depts/los/LEGISLATIONANDTREATIES/PDFFILES/PNG\_1977\_Act7.pdf}}. Analysis of the trajectory suggested an interstellar origin of the causative object CNEOS 2014-01-08: an arrival velocity relative to Earth in excess of $\sim 45~{\rm km~s^{-1}}$, and a vector tracked back to outside the plane of the ecliptic~\citep{SL22a}. The object's speed relative to the Local Standard of Rest of the Milky-Way galaxy, $\sim 60~{\rm km~s^{-1}}$, was higher than 95\% of the stars in the Sun’s vicinity. 

In 2022 the US Space Command issued a formal letter to NASA certifying a 99.999\% likelihood that the object was interstellar in origin~\footnote{\href{https://lweb.cfa.harvard.edu/~loeb/DoD.pdf}{https://lweb.cfa.harvard.edu/\textasciitilde loeb/DoD.pdf}}. Along with this letter, the US Government released the fireball lightcurve as measured by satellites~\footnote{\href{https://lweb.cfa.harvard.edu/~loeb/lightcurve.pdf}{https://lweb.cfa.harvard.edu/\textasciitilde loeb/lightcurve.pdf}}, which showed three flares separated by a tenth of a second from each other. The bolide broke apart at an unusually low altitude of $\sim$17 km, corresponding to a ram pressure of $\sim 200$ MPa. This implied that the object was substantially stronger than any of the other 272 objects in the CNEOS catalog, including the $\sim$5\%-fraction of iron meteorites from the solar system~\citep{SL22b}. Calculations of the fireball light energy suggest that about 500 kg of material was ablated by the fireball and converted into ablation spherules with a small efficiency~\citep{TR22}. The fireball path was localized to a 1 km-wide strip based on the delay in arrival time of the direct and reflected sound waves to a seismometer located on Manus Island~\citep{2023arXiv230307357S}. 

\section{Material Acquisition}

\begin{figure*}[!h]
     \centering
     \begin{subfigure}[t]{0.45\textwidth}
         \centering
\includegraphics[width=\textwidth]{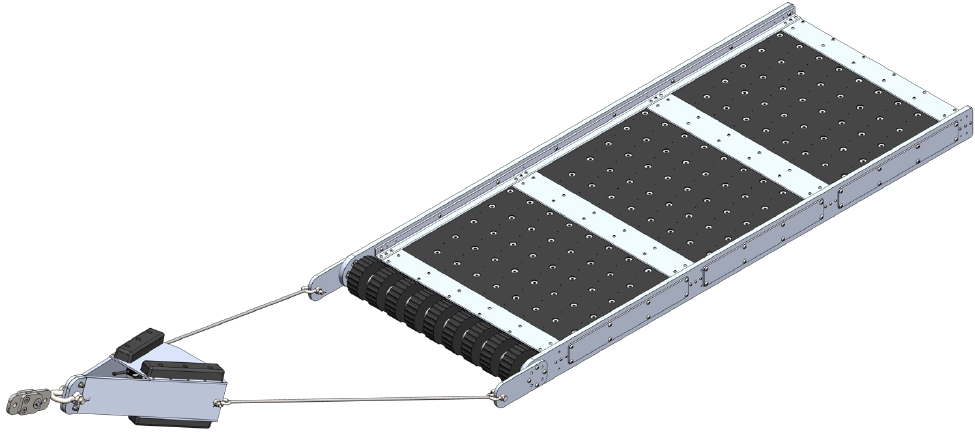}
\caption{Isometric view of the magnetic sled design.}
         \label{fig:slediso}
     \end{subfigure}
     \hfill
     \begin{subfigure}[t]{0.45\textwidth}
         \centering
\includegraphics[width=\textwidth]{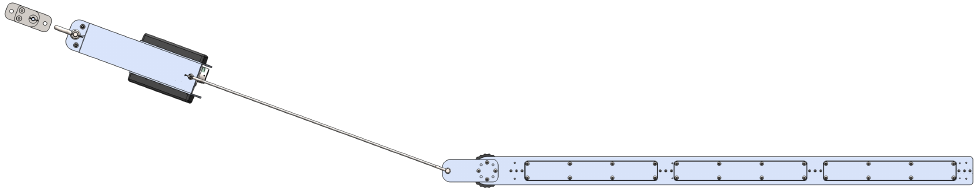}
\caption{Side view of the sled when placed on the ocean floor.}
         \label{fig:sledside}
     \end{subfigure}
     \caption{Magnetic sled design. The 200-kg, $1$ by $2$ meter sled was covered with an array of 300 neodymium magnets on both sides and equipped with video cameras in the metal tow-halter ahead of it, which was anchored by a synthetic cable to a winch on the ship, the M/V Silver Star.}
    \label{fig:sled}
\end{figure*}

An international expedition was organized and funded by a private contribution to search for remnants of the bolide, labeled hereafter IM1. The expedition was mounted from Port Moresby, Paupa New Guinea (PNG), and utilized a 40-meter catamaran workboat, the M/V Silver Star. A 200-kg sled was used with 300 neodymium magnets mounted on both of its sides and video cameras mounted on the tow-halter (see Figure~\ref{fig:sled}). After several experimental runs, the sled was observed to be `kiting' above the sea floor. To mitigate the kiting effect, 50 kg of lead were added to the sled.
Furthermore, the presence of heterogeneous surface currents and wind forces necessitated complex navigation strategies due to their effect on the vessel's maneuvering. Therefore, we systematically analyzed the seafloor current, enabling the alignment of the sled's trajectory with the prevailing current.

Following this, the sled was towed for an average of $\sim8$ hours per run, retrieved and the material was then processed. This process was repeated 26 times over a period of fourteen days. Several of these `run-tracks' were obtained from areas beyond the predefined target zone, acting as controls for the experiment.
We estimate that approximately $0.26~{\rm km^2}$ were sampled in the target area. 


\section{Sample Processing}
When the material collected from each run of the sled was brought aboard, it was examined carefully, with larger samples (mostly rusted iron) captured in vials for further analysis using X-ray fluorescence (XRF). 
The fine material collected on the neodymium magnets was then extracted and brought in a wet slurry up to a laboratory set up on the bridge of the vessel for further examination. 

There, an initial wet-magnetic separation took place; this was necessary because plankton and foraminifera were commonly entrained in the material attracted to the magnets. Subsequently, both magnetic and non-magnetic separations were processed through sieves and dried. Later an additional dry magnetic separation took place, and these secondary results were scanned with one of three digital microscopes aboard the ship. It was relatively easy to see and extract spherules above a millimeter in diameter with tweezers, but increasingly difficult to isolate below about 100 microns. A more detailed description of the sample-processing scheme is included in Appendix~\ref{app_a}. 

\begin{figure*}[t]
\includegraphics[width=12cm]{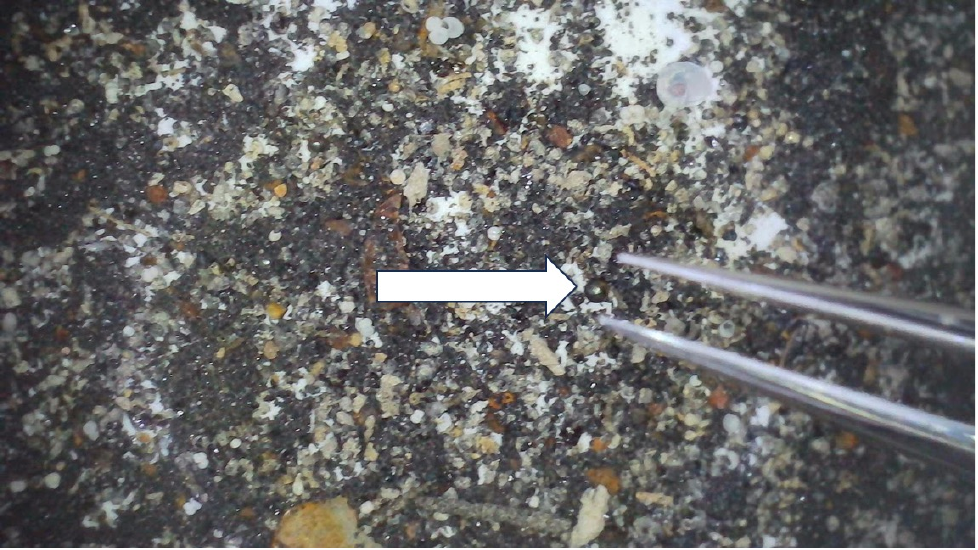}
\caption{Collected material from the magnetic sled at IM1's site, showing $\sim 0.4$mm diameter Fe-rich spherule (white arrow) amongst a background of shell hash and other debris}
\label{fig:spherule}
\end{figure*}

Initial X-ray Fluorescence (XRF) measurements of a few large spherules was attempted while on-board the research vessel using the Bruker CTX. The X-ray beam spot was several times larger than the samples, making backscatter and low absorption challenging. Despite this, reasonable XRF results were obtained for IM1 target area and background spherules.  

\begin{table}
\centering
\tiny
\begin{tabular}{|c|c|c|c|c|c|c|c|c|c|c|c|c|c|}
\toprule
\multicolumn{2}{|c|}{} & \multicolumn{11}{c|}{Radius (Microns) per spherule} & \multicolumn{1}{c|}{} \\
\hline
Run & Date (June/2023) &  &  &  &  &  &  &  &  &  &  &  & Count\\
\hline
1 & 14 & 0 & 0 & 0 & 0 & 0 & 0 & 0 & 0 & 0 & 0 & 0 & 0 \\
2 & 15 & 0 & 0 & 0 & 0 & 0 & 0 & 0 & 0 & 0 & 0 & 0 & 0 \\
3 & 16 & 0 & 0 & 0 & 0 & 0 & 0 & 0 & 0 & 0 & 0 & 0 & 0 \\
4 & 17 & 200 (SPH 11) & 350 (SPH 12) & 200 (SPH 13) & 150 (SPH 14) & 0 & 0 & 0 & 0 & 0 & 0 & 0 & 4 \\
5 & 18 & 75 & 0 & 0 & 0 & 0 & 0 & 0 & 0 & 0 & 0 & 0 & 1 \\
6 & 18 & 150 (SPH 1) & 0 & 0 & 0 & 0 & 0 & 0 & 0 & 0 & 0 & 0 & 1 \\
7 & 19 & 0 & 0 & 0 & 0 & 0 & 0 & 0 & 0 & 0 & 0 & 0 & 0 \\
8 & 20 & 200 (SPH 4) & 150 (SPH 5) & 400 (SPH 6) & 125 (SPH 7) & 200 (SPH 18) & 300 (SPH 19) & 100 (SPH 20) & 150 (SPH 21) & 150 (SPH 22) & 500 (SPH 23) & 200 (SPH 24) & 11 \\
9 & 20 & 400 (SPH 2) & 100 (SPH 25) & 400 (SPH 28) & 0 & 0 & 0 & 0 & 0 & 0 & 0 & 0 & 3 \\
10 & 21 & 0 & 0 & 0 & 0 & 0 & 0 & 0 & 0 & 0 & 0 & 0 & 0 \\
11 & 21 & 0 & 0 & 0 & 0 & 0 & 0 & 0 & 0 & 0 & 0 & 0 & 0 \\
12 & 22 & 150 (SPH 3) & 150 (SPH 10) & 400 (SPH 26) & 0 & 0 & 0 & 0 & 0 & 0 & 0 & 0 & 3 \\
13 & 23 & 550 (SPH 8) & 250 (SPH 29) & 0 & 0 & 0 & 0 & 0 & 0 & 0 & 0 & 0 & 2 \\
14 & 24 & 650 (SPH 1) & 0 & 0 & 0 & 0 & 0 & 0 & 0 & 0 & 0 & 0 & 1 \\
15 & 24 & 110 (SPH 15) & 200 (SPH 17) & 150 (SPH 30) & 0 & 0 & 0 & 0 & 0 & 0 & 0 & 0 & 3 \\
16 & 24 & 400 (SPH 9) & 150 (SPH 16) & 0 & 0 & 0 & 0 & 0 & 0 & 0 & 0 & 0 & 2 \\
17 & 25 & 250 (SPH 31) & 0 & 0 & 0 & 0 & 0 & 0 & 0 & 0 & 0 & 0 & 1 \\
18 & 25 & 0 & 0 & 0 & 0 & 0 & 0 & 0 & 0 & 0 & 0 & 0 & 0 \\
19 & 26 & 180 (SPH 33) & 150 (SPH 34) & 150 (SPH 35) & 210 (SPH 36) & 150 (SPH 37) & 250 (SPH 38) & 200 (SPH 39) & 150 (SPH 41) & 0 & 0 & 0 & 8 \\
20 & 26 & 270 (SPH 32) & 0 & 0 & 0 & 0 & 0 & 0 & 0 & 0 & 0 & 0 & 1 \\
21 & 26 & 200 (SPH 44) & 150 (SPH 45) & 200 (SPH 46) & 130 (SPH 47) & 0 & 0 & 0 & 0 & 0 & 0 & 0 & 4 \\
22 & 26 & 500 (SPH 40) & 110 (SPH 42) & 110 (SPH 43) & 0 & 0 & 0 & 0 & 0 & 0 & 0 & 0 & 3 \\
23 & 27 & 0 & 0 & 0 & 0 & 0 & 0 & 0 & 0 & 0 & 0 & 0 & 0 \\
24 & 27 & 200 (SPH 48) & 150 (SPH 49) & 200 (SPH 50) & 0 & 0 & 0 & 0 & 0 & 0 & 0 & 0 & 3 \\
\midrule
\multicolumn{12}{|c}{} & \multicolumn{2}{c|}{Total spherules: 51} \\
\bottomrule
\end{tabular}
\smallskip
\caption{Radii of the spherules (in microns) found in different runs on different dates in June 2023 while aboard the M/V Silver Star. Runs 25 and 26 on June 28 were not analyzed because of shortage of time at the end of the expedition.}
\label{tab:spherule_log_ship}
\end{table}
\begin{table}
    \centering
    \begin{tabular}{cccc}
        \toprule
        Run \# & \# of spherules & Avg. radius (microns) & Median radius (microns) \\
        \midrule
        4 & 25 & 168.8 & 150 \\
        5 & 1 & 75 & 75 \\
        6 & 1 & 150 & 150 \\
        7 & 2 & 225 & 225 \\
        8 & 44 & 188.2 & 155 \\
        9 & 17 & 199.4 & 150 \\
        10 & 3 & 150 & 140 \\
        11 & 26 & 207.3 & 170 \\
        12 & 86 & 169.5 & 150 \\
        13 & 102 & 183 & 160 \\
        14 & 57 & 194.4 & 170 \\
        15 & 24 & 197.9 & 160 \\
        16 & 23 & 192.6 & 170 \\
        17 & 59 & 314.8 & 291 \\
        18 & 6 & 161.7 & 165 \\
        19 & 43 & 209.6 & 180 \\
        20 & 33 & 191.5 & 170 \\
        21 & 4 & 170 & 175 \\
        22 & 46 & 220.2 & 165 \\
        23 & 2 & 130 & 130 \\
        24 & 18 & 201.1 & 200 \\
        \midrule
        Total: & 622 & & \\
        \bottomrule
    \end{tabular}
    \smallskip
    \caption{Statistical properties of spherules found at Harvard University after the expedition. }
    \label{tab:spherule_log_sophie}
\end{table}

\begin{figure*}
\includegraphics[width=14cm]{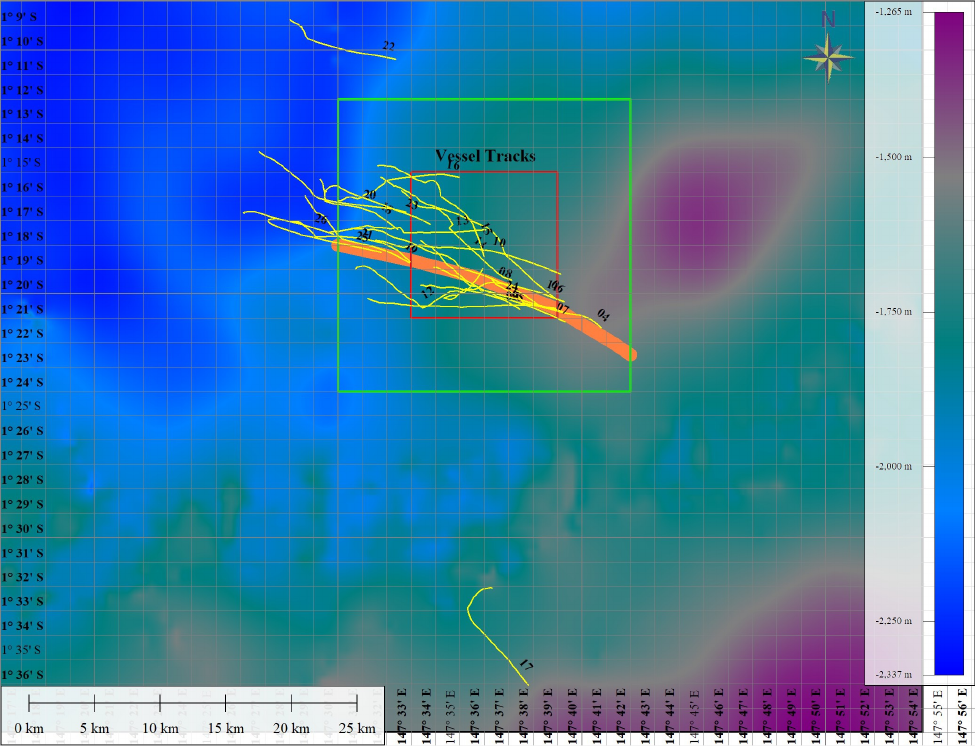}
\caption{List of ship track numbers around the expected path of IM1 based on seismometer data~\citep{2023arXiv230307357S} from Manus Island (orange strip). Background colors indicate ocean depth (with scale on the right). Latitude and longitude are marked in degrees and decimal minutes.  At one degree south latitude, one minute of latitude or longitude equals one nautical mile which is 1.852 km. The red box, measuring 11.112 km on a side, marks the uncertainty in the Department of Defence (DoD) localization of IM1's fireball, and the green box marks twice that size.}
\end{figure*}

\section{Spherule Count}
The spherule count in runs 1-24 is shown in Tables~\ref{tab:spherule_log_ship} 
and \ref{tab:spherule_log_sophie}
(with runs 25 \& 26 yet to be processed). We use the notation of ``IS" (as a short for ``Interstellar") to label specific run numbers and SPH to label specific spherule numbers based on the time they were discovered through the microscopes on the ship.  The average and median radii of spherules discovered later at the Harvard laboratory from the different runs are shown in Table~\ref{tab:spherule_log_sophie}.

\section{Spherule Location and Mass Distribution Analysis}

The synthetic cable from the ship to the sled was typically $\sim$5~km long and the ocean depth is on average $\sim$2~km in the search region. Only the vessel GPS coordinates were recorded, not the sled coordinates. Hence, we assign a cross-track error to the ship position to account for the uncertainty on the sled position. Assuming a deviation of the sled track of up to 30~degrees on either side of the ship track, a geometric calculation yields a cross-track error estimate of 2.29~km. We assign to each GPS record a 2.29~km error disc. The union of all discs for a given run constitute the area probed during the run.

The spherule total count $n_i$ for a given run $i$ is weighted by the total mass in grams $m_i$ of magnetic material analyzed from this run, providing the spherule yield parameter, $\rho_i$:
\begin{equation}
\rho_i = \frac{n_i}{m_i}.
\end{equation}

Assuming a uniform mass distribution of background magnetic material (mostly volcanic ash) in the sampled area, the amount of magnetic material collected and measured is representative of the total amount of material $M_i$ collected by the sled during run $i$: $m_i \propto M_i$. Material collected by the sled for a given run can vary due to external factors such as ocean currents, ship speed, sled angle, all of which affect the time spent on the ocean floor. Hence we represent the distribution of spherule counts weighted by the mass of magnetic material analyzed. Spherules were counted in the map only if their diameter was larger than 100~$\mu$m. 

We divided the sampled regions into a grid to make a heatmap. The contribution of a run (surface $\mathcal{R}_i$) to a given square pixel of the map's grid (surface $\mathcal{A}_{x, y}$) in the heatmap is proportionally weighted by the area overlap of the run and the square. If multiple runs contributed to the same pixel, we assigned the average spherule density to that pixel. 

We label as $\mathcal{I}$ the set of run numbers that overlap with a given pixel in row $x$, column $y$ of the grid. $|\mathcal{I}|$ is the cardinality of this set. The estimated spherule density $\xi_{x, y}$ in this pixel is computed as follows:
\begin{equation}
\xi_{x, y} = \frac{1}{|\mathcal{I}|}\sum_{i \in \mathcal{I}} \frac{|\mathcal{A}_{x, y} \cap \mathcal{R}_i|}{|\mathcal{A}_{x, y}|} \rho_i .
\end{equation}

\begin{figure}[h]
    \centering
    \includegraphics[width=7cm]{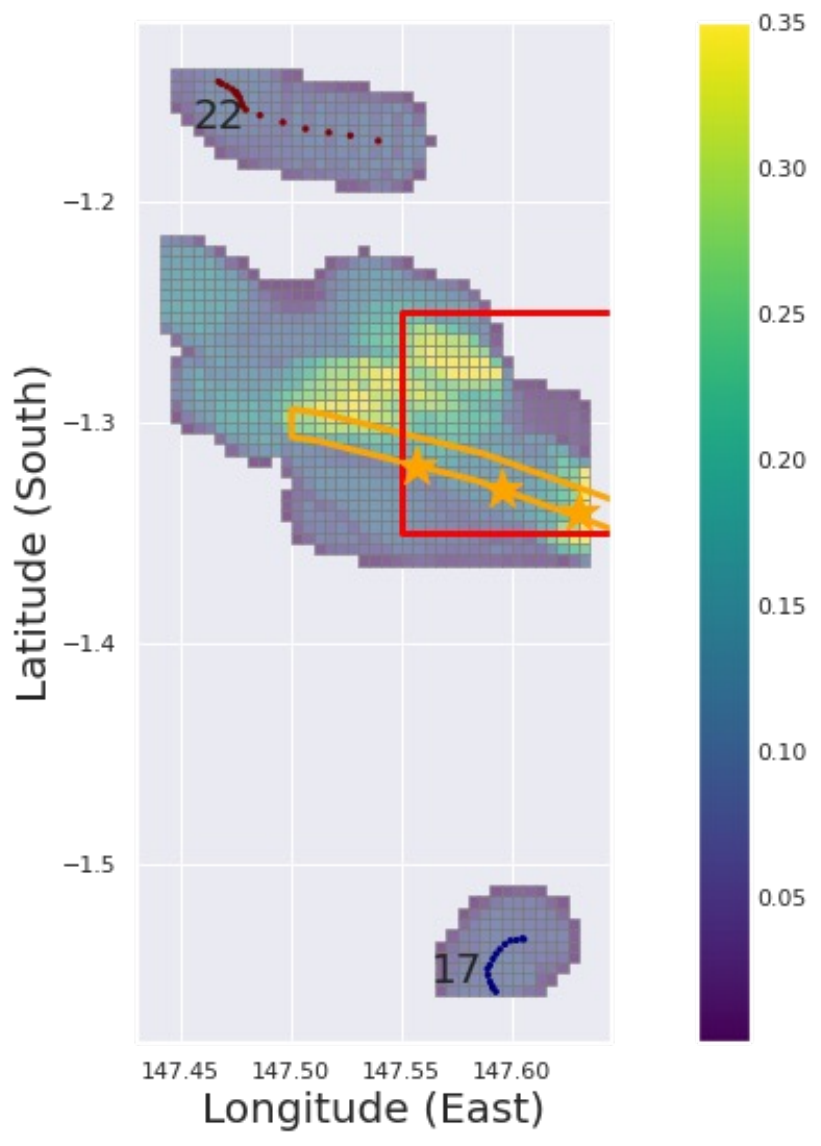}
    \caption{Heatmap of spherule density (count per mass of material analyzed in grams). Assuming that the first flare of the fireball lightcurve was located at the start of run 4, we placed three stars for the locations of the three flares. The color-bar maximum is clipped at 0.35 in this visualization. Each colored pixel in the heatmap is 0.555~km on a side. The negative latitudes stand for southern hemisphere. Runs 22 and 17 (control regions) are also labeled.}
    \label{fig:heatmap}
\end{figure}

\begin{figure}[h]
    \centering
    \includegraphics[width=12cm]{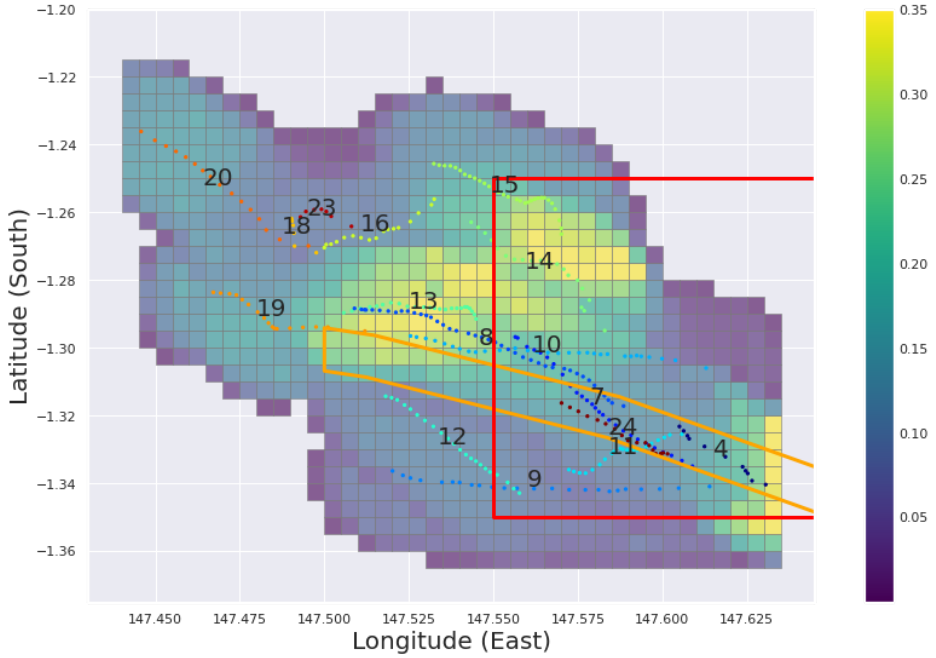}
    \caption{Heatmap of spherule density (count per mass of material analyzed in grams), zoomed onto the region sampled around the predicted IM1 path (orange box) and the DoD error region (red box). For reference, the dots represent the GPS recordings of the ship track in different numbered runs. The color-bar maximum is clipped at 0.35 in this visualization. Each pixel in this heatmap is 0.555~km on a side.}
    \label{fig:heatmap_zoom}
\end{figure}

Figure \ref{fig:heatmap} shows the resulting heatmap. Each colored pixel is 0.005 degrees (about 0.555~km) on a side. The yellow regions have a spherule density two times larger than the surrounding regions in blue. Figure \ref{fig:heatmap_zoom} is a zoomed version of the map onto the region close to the expected path of IM1. 


The high-yield (yellow) spherule regions occupy less than half of the surveyed area, allowing a fair comparison of the composition of IM1's spherules to background spherules. 
Given that the highest-yield (yellow) regions in the heat map account for roughly twice the background yield owing to the excess spherules added by IM1, we expect to find a significant numbers of IM1's spherules relative to background spherules in the highest-yield regions. As shown quantitatively in the next section, our composition analysis is consistent with that inference, suggesting that a significant fraction of the spherules in the yellow regions are of interstellar origin.

\section{Spherule Samples and Location}

The retrieval of cosmic spherules from meteor sites has a long history, with related morphology and composition analyses linking them to various components of the solar system~\citep{1979LPIBrownlee,maurette1991collection,1991MeticTaylor,1994MeticXue,brownlee1997leonard,herzog1999isotopic,taylor2000numbers,engrand2005isotopic,genge2008classification,vondrak2008chemical,wittke2013evidence,Folco2015,rudraswami2015evaluating,genge2017mineralogy}.

As demonstrated by the composition analysis described in the upcoming subsections, the excess ablation spherules collected from IM1's path are substantially different from the spherules collected in our control areas, which are expected to be solar system materials. 

Table~\ref{tab:spherule_log_four} shows the samples selected for elemental and isotopic measurements. The selected spherules range in diameter from 0.25 to 1.7 mm. They have been classified as I, S and D-types including some subclasses based on our measurements of their chemical compositions. The terminology and measurements will be explained in the upcoming subsections. We selected a substantial number of spherules from runs 4, 8, 13 and 14 along IM1’s path in Fig.~\ref{fig:heatmap_zoom}. Additional spherules were selected from run 22 and run 17, as they are not on IM1’s path and provide the background abundance patterns for spherules from the solar system.

\begin{longtable}[c]{ccccccc}
    \caption{Spherules analyzed for major and trace elements}\\
    \toprule
    Vial top label & Spherule name & Run \# & Type & Subclass & Calculated mass (mg) &Diameter (mm)\\
    \midrule
    \endfirsthead

        \toprule
    Vial top label & Spherule name & Run \# & Type & Subclass & Calculated mass (mg) &Diameter (mm)\\
    \midrule
\endhead
S10	& IS4B SPH1	&4	&D-Type	&{\bf BeLaU}	&0.327	&0.500\\
S21	&IS14 SPH1&	14&	D-Type&	{\bf BeLaU}&	1.727	&0.871\\
no label&	(6) IS14 SPH4&	14&	D-Type&	{\bf BeLaU}&	9.218&	1.522\\
no label&	(19) IS4 SPH8&	4& D-Type& {\bf BeLaU}&	0.847&	0.687\\
no label&	(25) IS13 SPH5&	13&	D-Type&	{\bf BeLaU}&	2.087&	1.045\\
no label& IS14 SPH2&	14&	D-Type&	Low-La&	2.681&	1.008\\
no label&	IS14 SPH3&	14&	D-Type&	Low-La&	0.209&	0.431\\
no label&	IS17 SPH1&	17&	D-Type&	Low-La&	0.096&	0.332\\
no label&	(1) IS22 SPH1&	22&	D-Type&	Low-La&	9.979&	1.760\\
no label&	(3) IS13 SPH8&	13&	D-Type&	Low-La&	1.579&	0.845\\
no label&	(11) IS14 SPH11&	14&	D-Type&	Low-La&	0.250&	0.457\\
no label&	(13) IS17 SPH3&	17&	D-Type&	Low-La&	0.270&	0.528\\
no label&	(14) IS17 SPH4&	17&	D-Type&	Low-La&	0.583&	0.683\\
no label&	(15) IS17 SPH5&	17&	D-Type&	Low-La&	0.383&	0.593\\
no label&	(17) IS4 SPH6&	4&	D-Type&	Low-La&	1.151&	0.761\\
no label&	(18) IS4 SPH7&	4&	D-Type&	Low-La&	0.575&	0.604\\
no label&	(21) IS8 SPH3&	8&	D-Type&	Low-La&	0.822& 0.680\\
no label&	(28) IS13 SPH8b&	13&	D-Type&	Low-La&	4.058&	1.304\\
no label&	(30) IS13 SPH12&	13&	D-Type&	Low-La&	1.982&	0.912\\
no label&	(31) IS14 SPH5& 14	&D-Type&	Low-La&	1.688&	0.864\\
no label&	(32) IS14 SPH6&	14	&D-Type&	Low-La&	0.425&	0.546	\\			
no label&	IS13 SPH3&	13	&I-type	&Ni-rich	&0.146	&0.383\\
S11	&IS4C SPH2&	4	&I-type	&Ni-rich	&0.097&	0.333\\
no label&	(12) IS14 SPH14&	14	& I-type&	Ni-rich&	0.496&	0.575\\
no label&	(24) IS8 SPH6&	8&	I-type&	Ni-rich&	0.832&	0.682\\
S16&	IS12C SPH1&	12	&I-type&	Ni, Mn-poor&	0.902&	0.701\\
no label&	IS8 SPH1&	8&	I-type&	Ni-poor& 0.497&	0.575\\
no label&	IS8 SPH2&	8&	I-type&	Ni-poor&	0.117&	0.355\\
S7&	IS16A SPH1& 16	&I-type&	Ni-poor&	0.836&	0.684\\
S23&	IS19 SPH1&	19	&I-type	&Ni-poor&	0.116&	0.354\\
S29&IS19 SPH7&	19&	I-type&	Ni-poor&	0.157&	0.392\\
S31&	IS19 SPH8&	19&	I-type&	Ni-poor	&0.119&	0.357\\
S5&	IS8-SPHR&	8&	I-type&	Ni-poor&	0.246&	0.455\\
S19&	IS13B SPH2&	13&	I-type&	Ni-poor&	0.495&	0.574\\
no label&	(2) IS22 SPH4	&22&	I-type&	Ni-poor&	4.643&	1.211\\
no label&	(5) IS13 SPH9&	13&	I-type&	Ni-poor	&2.840&	1.028\\
no label&	(10) IS14 SPH10&14&I-type&	Ni-poor	&0.777&	0.667\\
no label&	(22) IS8 SPH4	&8&	I-type&	Ni-poor&	0.615&	0.617\\
no label&	(26) IS13 SPH6	&13&	I-type&	Ni-poor&	2.068&	0.925\\
no label&	IS4 SPH4&	4&	S-type&	Chondritic&	0.149&	0.434\\
no label&	IS4 SPH5&	4&	S-type&	Chondritic&	0.071&	0.338\\
no label&	IS13 SPH4&	13&	S-type&	Chondritic&	0.122&	0.405\\
no label&	IS17 SPH2&	17&	S-type&	Chondritic&	0.075&	0.345\\
S25&	IS19 SPH2&	19&	S-type&	Chondritic&	0.079&	0.351\\
S12&	IS4D SPH3&	4&	S-type&	Chondritic& 0.078&	0.349\\
S9&	IS4A SPHS&	4&	S-type& Chondritic	&0.156&	0.440\\
no label&	IS22 SPH2&	22&	S-type&	Chondritic&	0.027&	0.244\\
no label&	IS22 SPH3&	22&	S-type&	Chondritic&	0.029&	0.251\\
no label&	(4) IS13 SPH11&	13&	S-type&	Chondritic& 0.639&	0.704\\
no label&	(7) IS14 SPH7&	14&	S-type&	Chondritic&	2.793&	1.151\\
no label&	(8) IS14 SPH8&	14&	S-type&	Chondritic&	0.711&	0.730\\
no label&	(9) IS14 SPH9&	14&	S-type&	Chondritic&	0.737&	0.738\\
no label&	(16) IS22 SPH5&	22&	S-type&	Chondritic&	1.793&	0.993\\
no label&	(20) IS4 SPH9&	4&	S-type&	Chondritic&	0.391&	0.598\\
no label&	(23) IS8 SPH5&	8&	S-type&	Chondritic&	0.597&	0.688\\
no label&	(27) IS13 SPH7&	13&	S-type&	Chondritic&	0.806&	0.761\\
no label&	(29) IS13 SPH10&	13&	S-type&	Chondritic&	0.658&	0.711\\
\bottomrule
    
    \smallskip
    
    \label{tab:spherule_log_four}
    \end{longtable}

\section{Analytical Methods}

We studied the spherules in four laboratories at Harvard University, UC Berkeley, Bruker Corporation in Germany and the University of Technology in Papua New Guinea. The elemental and isotopic measurements reported here are primarily from the Harvard Laboratory~\footnote{\href{https://projects.iq.harvard.edu/geochemistry}{https://projects.iq.harvard.edu/geochemistry}}. Additional results are mentioned in Appendix~\ref{app_b} and more will be presented in future publications.

\subsection{Electron microprobe analysis} 

Major element spot analyses and imaging of the spherules were performed with the JEOL Model JXA 8230 Electron Probe Microanalyzer (EPMA) in the Cosmochemistry Laboratory at Harvard University. The instrument uses either tungsten pin or LaB$_6$ filaments to generate an electron beam. The instrument is a high resolution, highly stable scanning electron microscope (SEM) equipped with 4 wavelength-dispersive spectrometers (WD) and one dry energy-dispersive (ED) spectrometer. The combination of 4 wavelength dispersive X-ray spectrometers (WDS) and a JED-2300 dry energy dispersive X-ray spectrometer (EDS) can simultaneously analyze 4 elements by WDS and many elements by EDS. This enables for example major elements measurements by EDS, and trace elements by WDS to save time. The operation software allows simultaneous displaying of up to 4 images including a backscattered electron image (BEI), a secondary electron image (SEI), a cathode-luminescence image (CLI), and X-ray images. In addition to point, line and area analysis and BSE, SEI, CLI, and X-ray imaging, the EPMA is capable for acquiring detailed chemical maps of larger areas which can be converted to phase or concentration maps using the built-in software and exported as images or digital maps for an off-line processing. Some samples are mounted in epoxy while others are on carbon tape for the imaging and spot measurements of the spherules.

\subsection{Triple Quad elemental analysis}

Measurements of elemental abundances for major and trace elements were performed on the iCAP TQ quadrupole ICP-MS (ThermoFisher Scientific) in the Cosmochemistry Laboratory at Harvard University. USGS reference materials were thoroughly dissolved and diluted in a 2\% HNO$_3$ solution spiked with 2 ppb indium diluted to a factor of ~5000 to be used as standards. Spherules were prepared for mass spectrometry measurements by first individually digesting the samples in a mixture of concentrated HF-HNO$_3$-HCl at a 1:3:1 ratio at 120-140$^\circ$C overnight. The samples were subsequently dried down and then redissolved in a second acid mixture involving an aqua regia solution mixed with H$_2$O at a 3:2 ratio and heated to $120^\circ$C overnight. This dissolution was dried down for the second time and redissolved in a high-purity 2\% HNO$_3$ solution. A small aliquot (3\%) was drawn from this solution and further diluted for elemental analysis. To account for and to correct instrumental drift, the 2\% HNO$_3$ solution used for dilution was spiked with 2 ppb indium as an internal standard, prepared identically to the standard solutions. Measurements were performed in KED mode with He as a collision cell gas as recommended by the Reaction Finder function built in the iCAP TQ software, with the exception of Cr, which was measured in TQ mode with O$_2$ as a mass-shifted molecule. The prepared spherule solutions were measured as an unknown against a four-point calibration line consisting of a blank and three USGS standards: BCR-2 BHVO-2, and AGV-2. Calibration curves for individual elements were checked for linear intensity to concentration correlations for accurate measurements. Routine measurements of AGV-2 as an unknown on the iCAP TQ suggest fractional errors to be within 6\% using this method.

\subsection{Iron isotope analysis with the Collision Cell Multicollector ICP} 

Measurements of iron isotopes were performed with the Nu Sapphire multi-collector inductively coupled plasma mass spectrometer (MC-ICPMS) in the Cosmochemistry Laboratory at Harvard University. All Fe ion beams may be affected by molecular interferences from argide ions. Hydrogen and He were introduced into the collision-reaction cell of the instrument to reduce interferences from Ar-related species ($^{40}$Ar$^{14}$N+ on $^54$Fe, $^40$Ar$^{16}$O$^+$ on $^{56}$Fe and $^{40}$Ar$^{16}$O$^1$H$^{+}$ on $^{57}$Fe). Sample solutions were introduced into the instrument using an Elemental Scientific Apex Omega desolvating spray chamber with a 100 microliter per minute PFA nebulizer. Iron was purified from the sample matrix by passing an aliquot of solution once through BioRad AG1-X8 ion-exchange resin. Solution Fe purity was checked by Quadrupole MC-ICPMS, prior to isotopic analysis.  Iron isotope ratios ($^{56}$Fe/$^{54}$Fe, $^{57}$Fe/$^{54}$Fe) were analyzed by sample-standard bracketing relative to a lab standard.  Sample and standard solutions were prepared at concentrations of about 70 ppb.  All solutions were matched in concentration to within about 2\% or better to avoid significant concentration-related matrix effects.

Mass-dependent isotope values are reported as $\delta$-values. 
The $\delta^{56/54}$Fe and $\delta^{57/54}$Fe values are defined as the per mil deviations of the measured isotope ratios,  ($^{56}\text{Fe}/^{54}\text{Fe} )_{\text{meas}}$ and ($^{57}\text{Fe}/^{54}\text{Fe} )_{\text{meas}}$, relative to the same ratios in a standard (std):
\begin{equation}
\delta^{^{56/54}}\text{Fe} = \left( \frac{\left( \frac{^{56}\text{Fe}}{^{54}\text{Fe}} \right)_{\text{meas}}}{\left( \frac{^{56}\text{Fe}}{^{54}\text{Fe}} \right)_{\text{std}}} - 1 \right) \times 10^3, 
\end{equation}
\begin{equation}
\delta^{^{57/54}}\text{Fe} = \left( \frac{\left( \frac{^{57}\text{Fe}}{^{54}\text{Fe}} \right)_{\text{meas}}}{\left( \frac{^{57}\text{Fe}}{^{54}\text{Fe}} \right)_{\text{std}}} - 1 \right) \times 10^3.
\end{equation}
Analyses of the common standard NIST IRMM-14 were obtained relative to the lab standard, so that $\delta$-values relative to the lab standard can be easily converted and compared to literature values for iron isotopes relative to NIST IRMM-14.

In addition, the $^{57}$Fe/$^{54}$Fe ratio was corrected for mass fractionation using the $^{56}$Fe/$^{54}$Fe ratio.  The \(\varepsilon^{57/54}\text{Fe}\), 
defined as the part in \(10^4\) deviation of the fractionation corrected isotope ratios, \(\left( \frac{^{57}\text{Fe}}{^{54}\text{Fe}} \right)_{\text{meas}(N)}\), 
relative to the same ratio in a standard (std), is:
\begin{equation}
\varepsilon^{^{57/54}}\text{Fe} = \left( \frac{\left( \frac{^{57}\text{Fe}}{^{54}\text{Fe}} \right)_{\text{meas(N)}}}{\left( \frac{^{57}\text{Fe}}{^{54}\text{Fe}} \right)_{\text{std}}} - 1 \right) \times 10^4,
\end{equation}
Uncertainties reported in figures and tables are internal mass spectrometric errors, and the true uncertainties may be somewhat greater.  

\subsection{Imaging, morphology and chemical spot analyses of the spherules}

Spherules collected near IM1's path appear to be nested (see Appendix~\ref{app_b}), suggesting that liquid drops engulfed smaller ones which solidified earlier. The dendritic textures of these spherules suggests rapid cooling. 

\begin{figure*}[t]
\includegraphics[width=10cm]{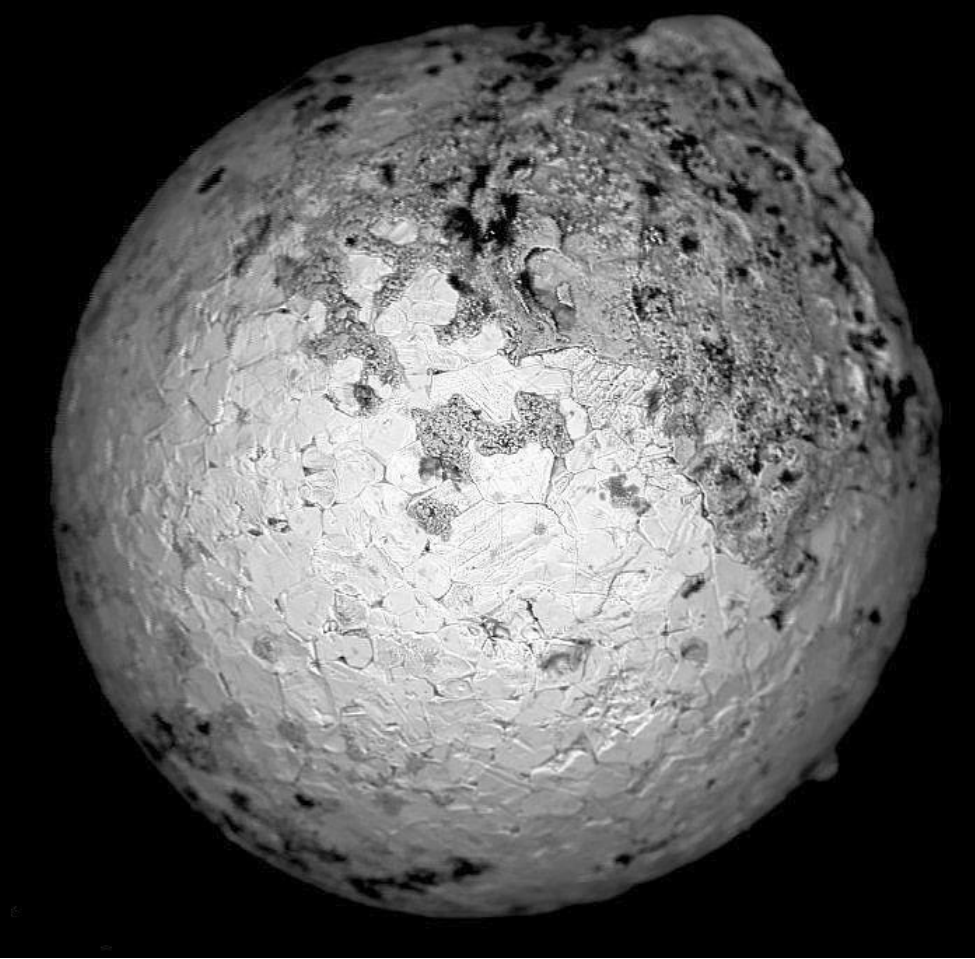}
\caption{Electron microprobe image of spherule IS16A SPH1 from run 16.}
\label{fig:soccerball}
\end{figure*}

\begin{figure*}
\includegraphics[width=10cm]{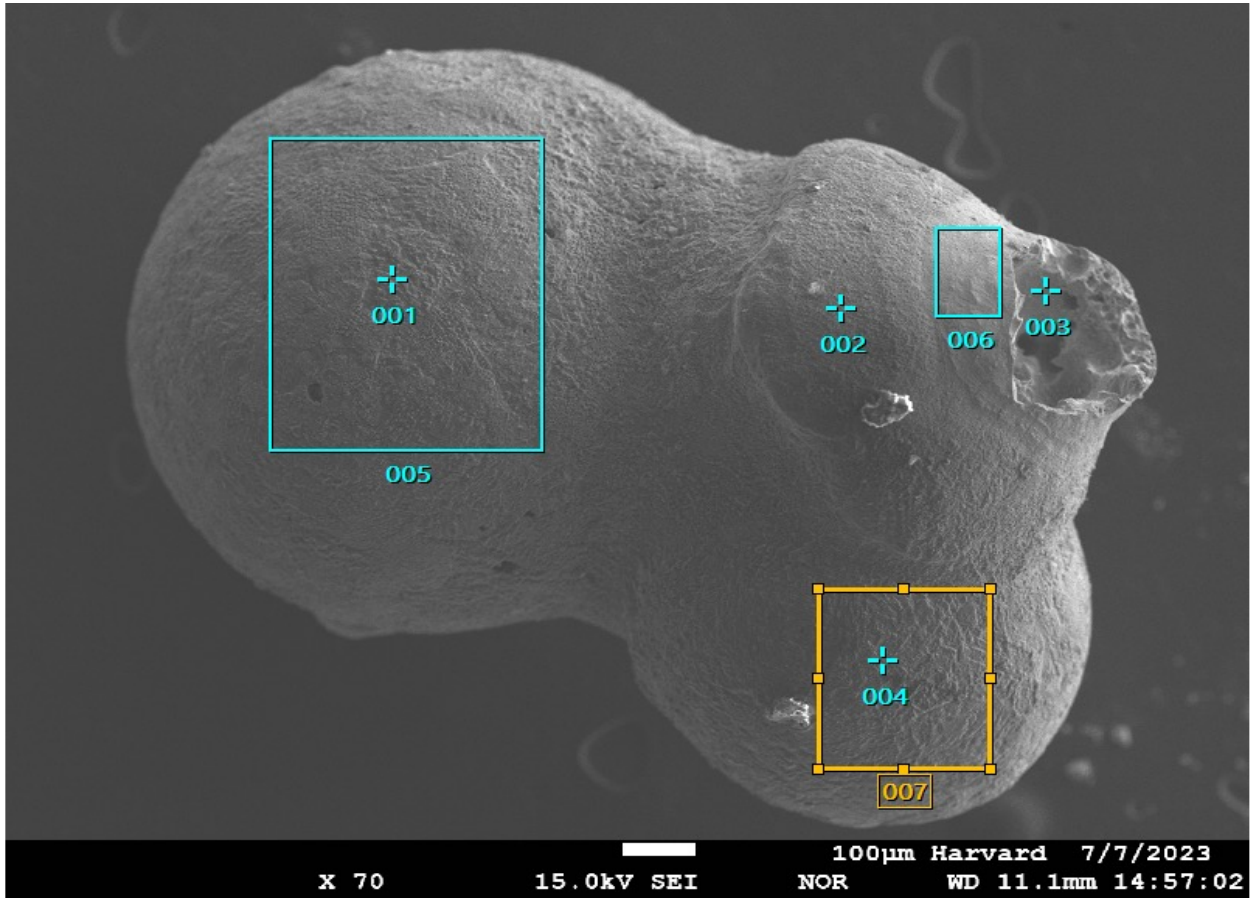}
\caption{Electron microprobe image of the spherule S21 (IS14-SPH1) from run 14 in the high yield region of IM1's path. The elemental abundances measured in different regions on the surface are displayed in Table \ref{fig:slide4}.}
\label{fig:slide3}
\end{figure*}

Figure~\ref{fig:soccerball} shows an electron microprobe image of a modest-size (0.7 mm in diameter) spherule IS16A SPH1 from run 16. An example of a large (1.3 mm in maximum diameter, 0.9 mm average) spherule in the high-yield (yellow) region along IM1's path is S21 (IS14-SPH1) from run 14. This lopsided spherule, shown in Figure~\ref{fig:slide3}, is a composite of three spherules that solidified shortly after merger but too late for the merger product to become spherical. The mass of S21 (1.7 mg) is about twice that of IS16A SPH1 (0.84 mg).

\begin{table}
  \centering
  \begin{tabular}{lccccccc}
    \multicolumn{1}{c}{} & \multicolumn{7}{c}{} \\
             Element & Pt\#1 & Pt\#2 & Pt\#3 & Pt\#4 & Pt\#5 & Pt\#6 & Pt\#7 \\
             & (weight \%) & (weight \%) & (weight \%) & (weight \%) & (weight \%) & (weight \%) & (weight \%) \\
    \hline
    C    & 0.64  &       & 0.7   & 1.5  & 2.1  & 5.48 & 2.66 \\
    O    & 13.95 & 28.9  &       & 32.35& 29.18& 18.66& 34.19\\
    Na   & 0.63  & 0.73  &       &      & 0.7  & 0.72 &      \\
    Mg   & 0.28  &       &       &      & 0.42 & 0.15 &      \\
    Al   & 2.93  & 3.48  &       & 1.15 & 2.45 & 1.44 & 1.18 \\
    Si   & 6.4   & 10.85 &       &      & 2.92 & 4.2  & 0.97 \\
    P    &       & 0.31  &       &      &      &      &      \\
    Cl   & 0.26  & 0.4   &       &      & 0.51 & 1.28 &      \\
    K    & 0.39  &       &       &      &      &      &      \\
    Ca   & 3.88  & 4.55  &       &      & 1.73 & 1.85 & 0.23 \\
    Cr   &       &       &       &      & 0.31 &      &      \\
    Fe   & 70.65 & 50.78 & 99.3  & 65.01& 59.69& 66.22& 60.77\\
    Total& 100   & 100   & 100   & 100  & 100  & 100  & 100  \\
    \hline
  \end{tabular}
  \caption{Electron microprobe data on the chemical composition in the different regions on the surface of S21, as labeled in Figure \ref{fig:slide3}.}
  \label{fig:slide4}
\end{table}

The existence of a triple-merger like S21 can be explained as a product of IM1's airburst. The total mass collected by spherules in our 26 runs is of order $\sim 1$g. Given the sled's width of 1m, the total surveyed area, $\sim 0.26~{\rm km^2}$, constitutes a fraction of $\sim 10^{-2}$ of IM1's strewn field, defined by the yellow regions in Figure~\ref{fig:heatmap_zoom}. This implies a total mass in IM1's spherules of order $\sim 100$g, as expected given that most of IM1's mass evaporated to undetectable particles (well below tens of $\mu$m in size) or gas~\citep{TR22}. Assigning a mass of $\sim 10^{-3}$g per spherule implies a total number of $\sim 10^5$ such spherules. Based on IM1's speed and fireball energy, the total mass ablated by IM1's fireball is $\sim 5\times 10^5$g corresponding to an object radius of $R\sim 50$cm ~\citep{SL22a}. The total number of spherules divided by the initial volume associated with $R$ yields an initial spherule number density of $n\sim 0.2~{\rm cm^{-3}}$ which gets diluted considerably as the material expands. For the characteristic diameter of a spherule, $\sim 1$mm, the geometric cross-section for spherule-spherule collisions is $\sigma \sim 3\times 10^{-2}~{\rm cm^2}$.  The resulting collision probability is $\tau\sim n\sigma R\sim (0.2~{\rm cm^{-3}})\times (3\times 10^{-2}~{\rm cm^2})\times (50~{\rm cm})\sim 0.3$, implying a likelihood of $\tau^2\sim 0.1$ for triple-spherule mergers such as S21. Mergers that occur inside a liquid envelope would result in a spherical shell with embedded sub-spherules inside of it, as shown in the images of Appendix~\ref{app_b}.

\subsection{Classification by elemental composition measurements of bulk spherules}

\begin{figure*}
\includegraphics[width=10cm]{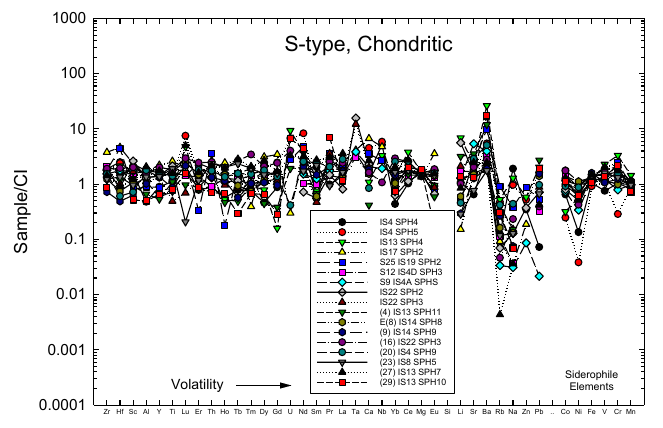}
\caption{Abundances of elements as a function of their volatility for S-type chondritic spherules from run 4, 8, 9, 12, 13, 14, 17, 19 and 22.}
\label{fig:volatility}
\end{figure*}

\begin{figure*}
     \centering
     \begin{subfigure}[b]{0.49\textwidth}
         \centering
\includegraphics[width=\textwidth]{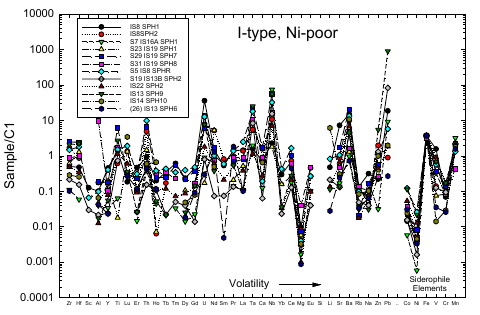}
\caption{}
         \label{fig:volatility_i_ni_poor}
     \end{subfigure}
     \hfill
     \begin{subfigure}[b]{0.49\textwidth}
         \centering
\includegraphics[width=\textwidth]{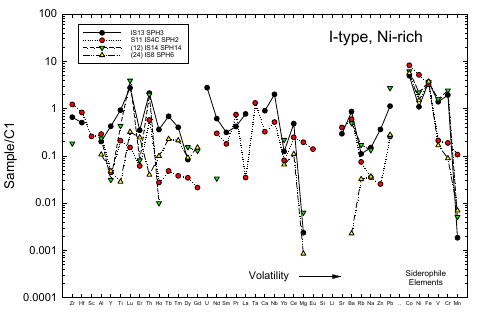}
\caption{}
         \label{fig:volatility_i_nimncr}
     \end{subfigure}
     \caption{Abundances of elements as function of their volatility in I-type spherules. (a) I-type, Ni-poor spherules from runs 8, 13, 14, 16, 19, and 22. (b) I-type, Ni-rich spherules from runs 4, 8, 13 and 14.}
    \label{fig:volatility_i}
\end{figure*}

\begin{figure*}
     \centering
     \begin{subfigure}[b]{0.49\textwidth}
         \centering
\includegraphics[width=\textwidth]{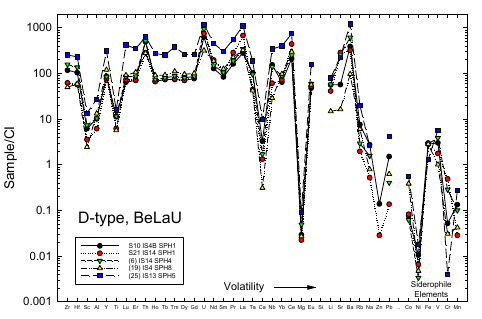}
\caption{}
         \label{fig:volatility_i_ni}
     \end{subfigure}
     \hfill
     \begin{subfigure}[b]{0.49\textwidth}
         \centering
\includegraphics[width=\textwidth]{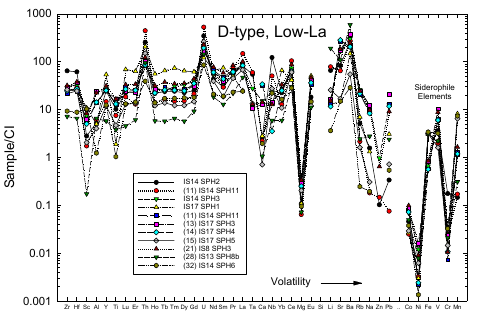}
\caption{}
         \label{fig:volatility_low_la}
     \end{subfigure}
\caption{Abundances of elements as function of their volatility for D-type spherules: (a) ``BeLaU"-type spherules from runs 4, 13 and 14, from the high-yield regions near IM1's path; and (b) D-type, Low-La spherules from runs 8, 13, 14 and 17.}
\label{fig:volatility_belau}
\end{figure*}

\begin{figure*}
\includegraphics[width=10cm]{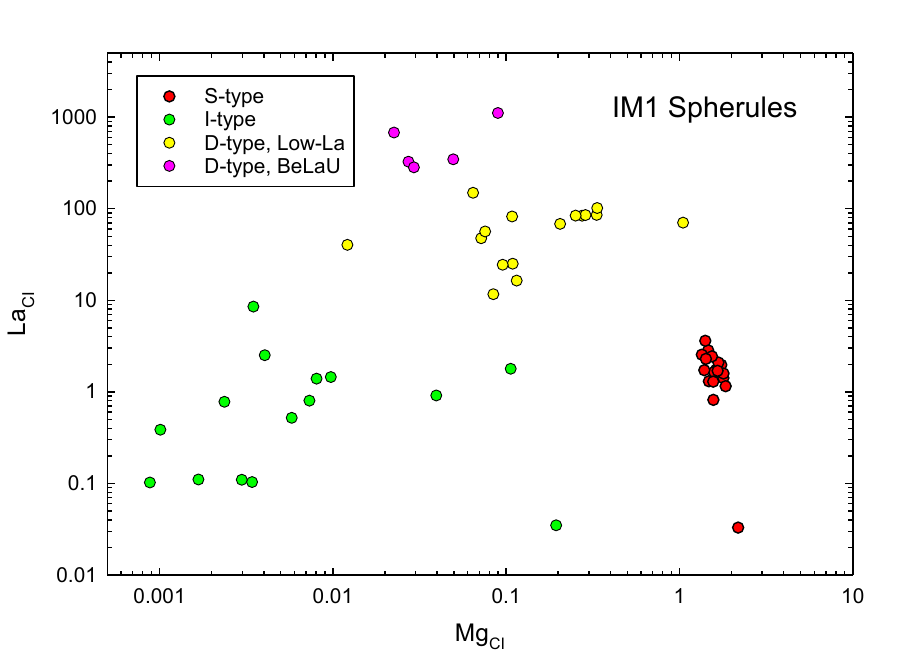}
\caption{
CI-normalized Mg and La for the 57 analyzed spherules. 
}
\label{fig:la_vs_mg}
\end{figure*}

We used Figure 8 and tables 2 and 3 in a review by~\citet{Folco2015} to classify the 57 spherules listed in Table 4. Our elemental data are provided in Appendix~\ref{sec:elemental_data}. The elements are plotted as a function of their volatility in Figures~\ref{fig:volatility}, \ref{fig:volatility_i} and \ref{fig:volatility_belau}, using the same elements as~\citet{Folco2015} in their Figure 8. This resulted in 18 spherules being classified as standard S-type chondritic spherules (Figure~\ref{fig:volatility}) and another 18 spherules being classified as standard I-type spherules (Figure~\ref{fig:volatility_i}). Four of the I-type spherules are Ni-rich, while the rest are Ni-poor. In addition, we identified 21 spherules that are clearly derived from material that has gone through planetary differentiation as they have many features, such as low Mg, that are typical of planetary differentiation (Figure~\ref{fig:volatility_belau}). They are all different from the differentiated spherule pattern described by~\citet{Folco2015}. We refer to these new types of differentiated spherules as D-type spherules. These have never been described in the cosmic spherule literature. There are two subtypes of the D-type differentiated spherules. We call the subtypes Low-La and ``BeLaU" and their distinct elemental patterns are shown in Figure~\ref{fig:volatility_belau}. The ``BeLaU" spherules have refractory lithophile element abundances that are 80 to 1,000 higher than in CI chondrites. The highly enriched ``BeLaU" spherules (enriched in Be, La, U and many other refractory lithophile elements) will be discussed in more detail in the next section. Figure~\ref{fig:la_vs_mg} shows the CI-normalized Mg and La for the 57 spherules. These two elements provide a relatively clear distinction between the spherule groups discussed here. 

A summary of spherule types found in individual runs is listed in Table~\ref{tab:spherule_type_summary}. This
shows that ``BeLaU"-type spherules are only found in runs 4, 13 and 14 which overlap with the high yield (yellow) IM1 regions in the heatmap of Fig.~\ref{fig:heatmap_zoom}. The lines of these three runs also stretch across low-yield regions which moderate their average yield of ``BeLaU" spherules. With less than a third of the total length of runs 4, 13 and 14 lying in high-yield (yellow) regions, the retrieval of 5 ``BeLaU"-type spherules out of a total of 34 spherules analyzed in these three runs, suggests that the high-yield (yellow) regions should have comparable numbers of ``BeLaU"-type and background spherules. This agrees with the enhancement by a factor of $\sim 2$ in the number of spherules per unit mass associated with the high-yield (yellow) regions along IM1's path, compared to the background (green-purple) in the heatmap of Fig.~\ref{fig:heatmap_zoom}. 

\begin{table}[ht]
  \centering
  \begin{tabular}{lcccccc}
    \toprule
    Run & Number of spherules & S-type & I-type & D-type (Low-La) & D-type (BeLaU) \\
    \midrule
    IS4   & 10                 & 5      & 1     & 2      & 2      \\
    IS8   & 7                  & 1      & 5     & 1      & 0      \\
    IS12  & 1                  & 0      & 1     & 0      & 0      \\
    IS13  & 12                 & 4      & 4     & 3      & 1      \\
    IS14  & 12                 & 3      & 2     & 5      & 2      \\
    IS16  & 1                  & 0      & 1     & 0      & 0      \\
    IS17  & 5                  & 1      & 0     & 4      & 0      \\
    IS19  & 4                  & 1      & 3     & 0      & 0      \\
    IS22  & 5                  & 3      & 1     & 1      & 0      \\
    Total & 57                 & 18     & 18    & 16     & 5      \\
    \bottomrule
  \end{tabular}
  \caption{Results summary for 57 spherules analyzed for major and trace element composition.
  }
  \label{tab:spherule_type_summary}
\end{table}


\subsection{Significance of the novel "BeLaU"-type spherules}

\begin{figure*}
\includegraphics[width=10cm]{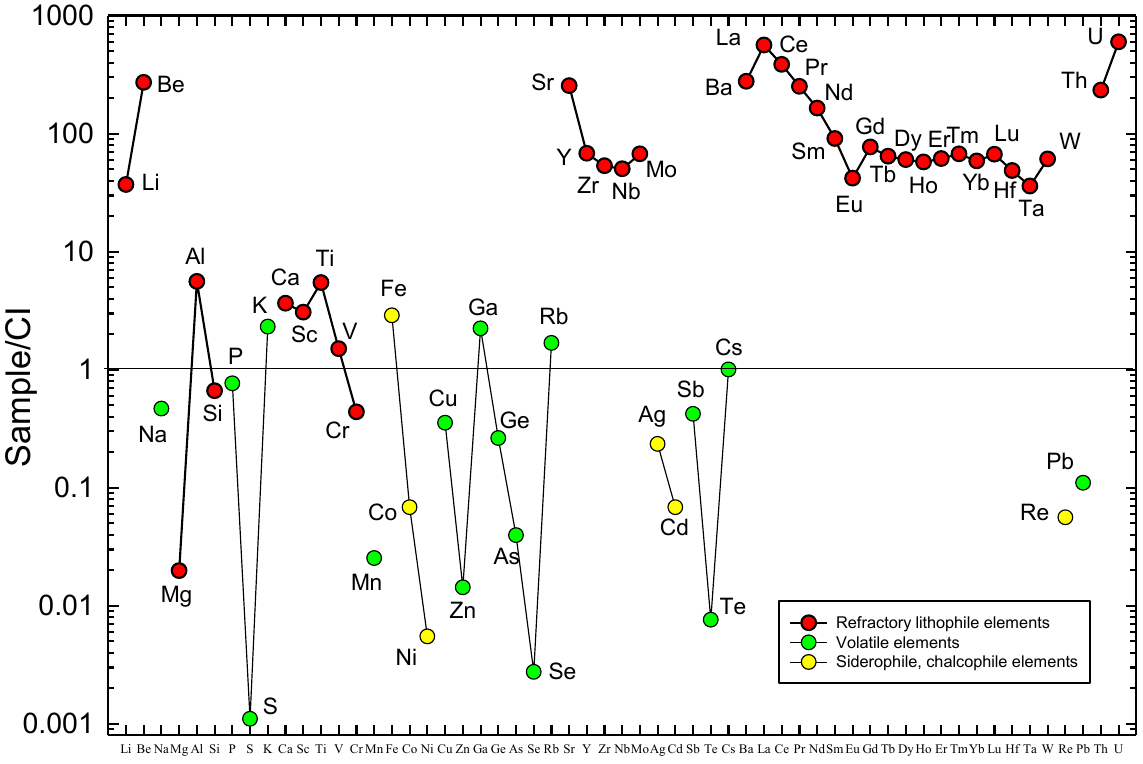}
\caption{The ``BeLaU" elemental abundances of the spherule S21 (normalized to CI chondrites) versus atomic number for 56 elements. The solar system standard of CI chondrites is represented by a value of unity on the plot.
}
\label{fig:slide13}
\end{figure*}

\begin{figure*}
\includegraphics[width=10cm]{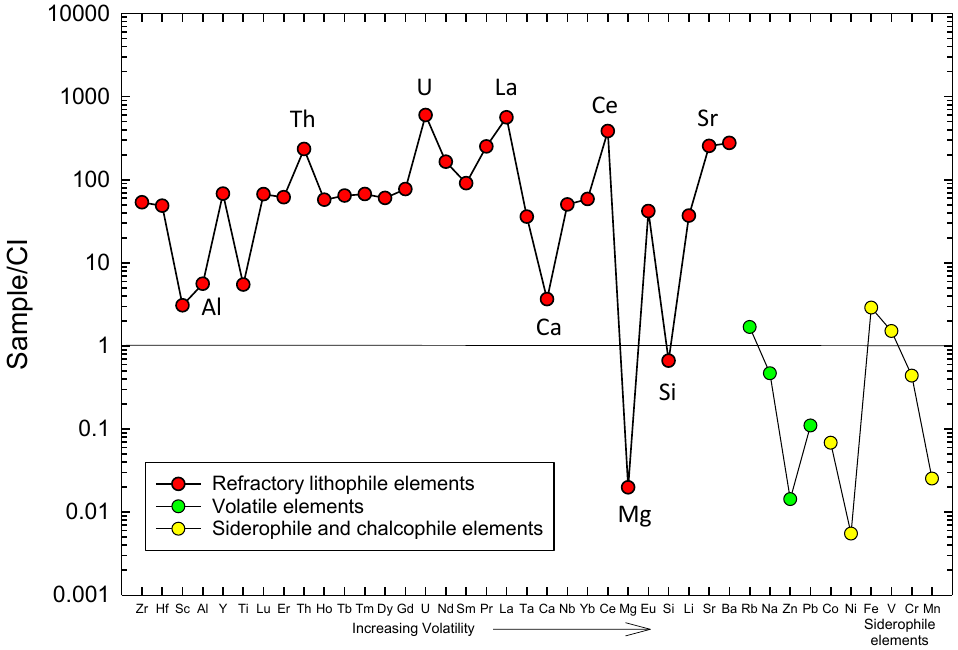}
\caption{The abundance pattern of elements in spherule S21 versus volatility.}
\label{fig:slide14}
\end{figure*}

\begin{figure*}
\includegraphics[width=10cm]{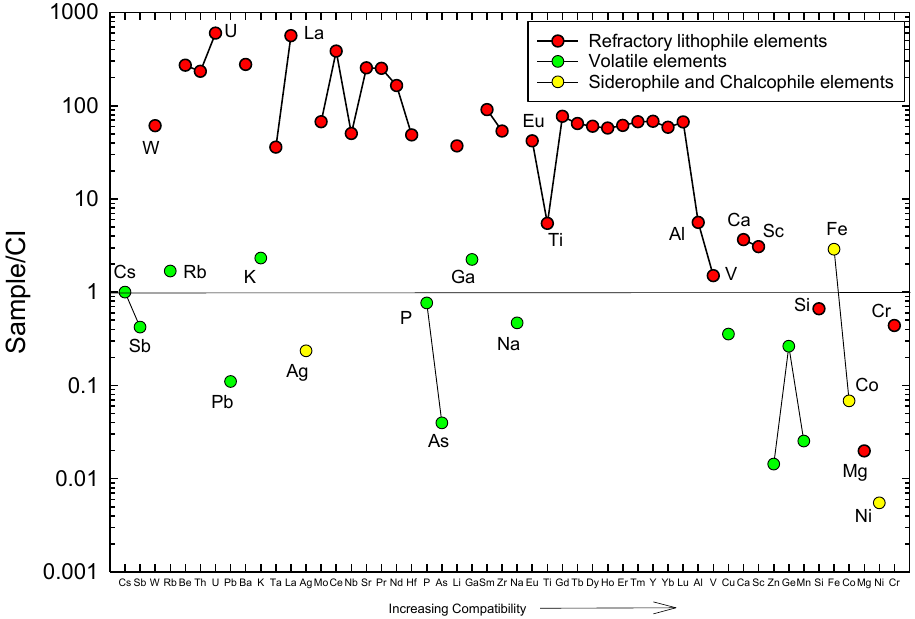}
\caption{Abundances versus compatibility (see text) for S21. Elements ordered with increasing compatibility towards the right. }
\label{fig:slide15}
\end{figure*}

The spherules with enrichment of beryllium (Be), lanthanum (La) and uranium (U), labeled ``BeLaU", appear to have an exotic composition different from other solar system materials. The results for the ``BeLaU" spherule S21 are displayed in Figures~\ref{fig:slide13},~\ref{fig:slide14} and~\ref{fig:slide15}. We use this spherule to point out some of the unique feature of the ``BeLaU" elemental composition. 

A plot of the elemental abundances of the spherule S21 (normalized to CI chondrites) as a function of atomic number for 56 elements is shown in Figure~\ref{fig:slide13}. Across the diagram the peak abundances are for Be, La and U, hence the name ``BeLaU". The abundance pattern of S21 implies derivation from a planetary crust, highly enriched in refractory lithophile elements (red dots). The volatile element abundances (green dots) are very low, suggesting either derivation from an extremely volatile-depleted planet or evaporative loss during passage through the Earth’s atmosphere. The very low content of refractory siderophile elements with affinity to iron (Re) suggest a source planet with an Fe core. A plot of the abundance pattern of elements in spherule S21 as a function of the volatility sequence used by~\citet{Folco2015}  is shown in Figure~\ref{fig:slide14}. Since there are strong indications that the spherules are derived from a differentiated planet, the data are also plotted in Figure~\ref{fig:slide15} as a function of an igneous compatibility sequence. Compatibility is a geochemical parameter measuring how readily a particular element substitutes for a major element in mantle source minerals during melting to produce magma. It also roughly represents the sequence of enrichments of elements in a crystallizing magma.

\begin{figure*}
\includegraphics[width=10cm]{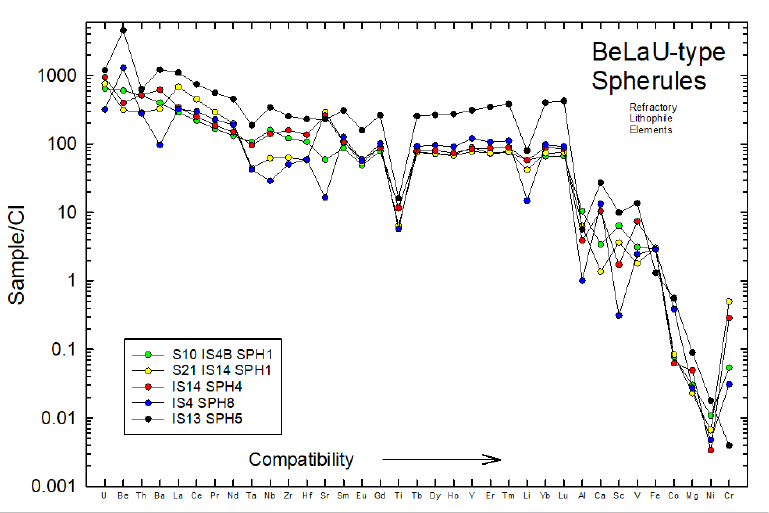}
\caption{Abundances of elements as function of their compatibility for all ``BeLaU"-type spherules. They are all from runs 4, 13 and 14, the high-yield (yellow) regions near IM1’s path in Fig.~\ref{fig:heatmap_zoom}.}
\label{fig:slide16}
\end{figure*}

The abundances of elements as function of their compatibility for all 5 ``BeLaU"-type spherules are shown in Figure~\ref{fig:slide16}. They are all from runs 4, 13 and 14, the high-yield (yellow) regions near IM1’s path. The ``BeLaU" spherules' variations in the abundances of trace elements relative to CI chondrites are higher by 1-3 orders of magnitude compared to cosmic spherules from the solar system reviewed by~\citet{Folco2015}. This ``BeLaU" abundance pattern found in IM1’s spherules could have possibly originated from a highly differentiated planetary magma ocean. 

\begin{figure*}
\includegraphics[width=10cm]{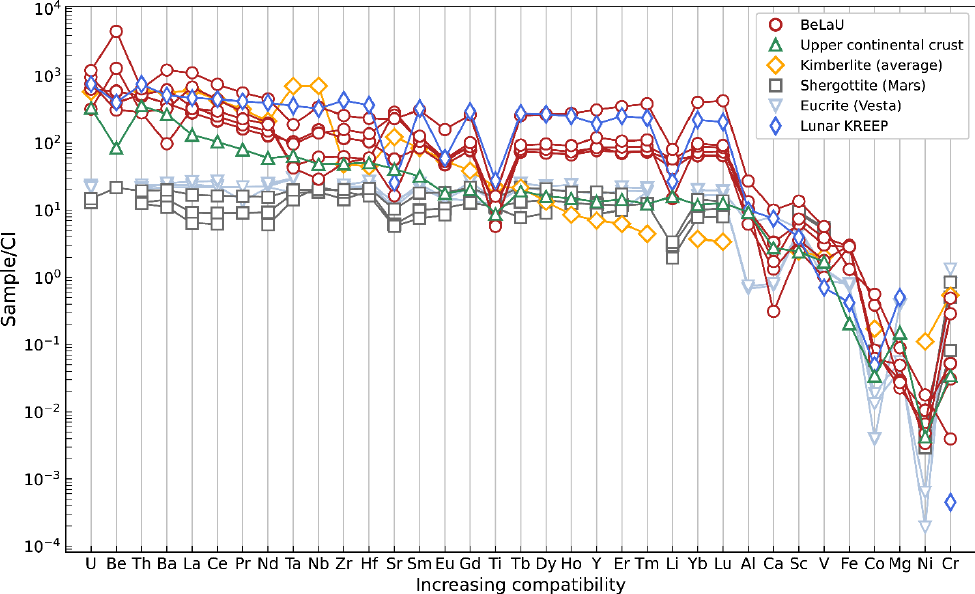}
\caption{The ``BeLaU" element abundance pattern has features that make it distinct from enriched sources on solar system bodies. Elements are ordered with increasing compatibility. Data source: Earth’s upper continental crust~\citep{rudnick2014composition}; kimberlites~\citep{giuliani2020kimberlite}; lunar KREEP~\citep{warren1989kreep}; shergottites [Shergotty, Zagami, and Los Angeles meteorites]~\citep{lodders1998survey,jambon2002basaltic}; eucrites [Bouvante, Pomozdino, and Stannern meteorites]~\citep{Kitts}.
}
\label{fig:slide17}
\end{figure*}

The ``BeLaU" element patterns are compared in Figure~\ref{fig:slide17} with the highly enriched sources of differentiated bodies in the solar system, including Earth’s upper continental crust~\citep{rudnick2014composition} and kimberlites~\citep{giuliani2020kimberlite}, lunar magma ocean residual liquid (KREEP)~\citep{warren1989kreep}, Shergottites from Mars~\citep{lodders1998survey, jambon2002basaltic} and eucrites from Vesta~\citep{kitts1998survey}. Figure~\ref{fig:slide17} presents the comparison of refractory lithophiles that are not easily lost by evaporation or altered by fluids. Shergottites and eucrites are markedly distinct from the ``BeLaU" samples due to the systematically lower incompatible element enrichments. The upper continental crust is also overall depleted in incompatible elements compared to ``BeLaU". Additionally, ``BeLaU" samples have prominent negative anomalies of Ti, Li, higher Lu/Al, and variable Be and Sr enrichments that do not match the smooth pattern of the upper continental crust. Kimberlite is also remarkably distinguished from the ``BeLaU" pattern in their Ta and Nb positive anomalies and the strong heavy rare earth element depletion. Lastly, despite the resembling overall element enrichments, the lunar KREEP displays pronounced differences in light rare earth element enrichment and strong Sr, Eu, and Cr negative anomalies from ``BeLaU" that distinguish the two groups. We conclude that the ``BeLaU" samples possibly reflect an extremely evolved composition from a planetary magma ocean, but not the known bodies within the solar system.

Table~\ref{tab:spherule_log_four} shows that ``BeLaU"-type spherules were retrieved from runs 4, 13 and 14  in the high-yield (yellow) regions near IM1's path in Fig.~\ref{fig:heatmap_zoom}. Additional spherules from run 22 and run 17 in control regions exhibited background abundance patterns from the solar system.

\subsection{Iron isotope measurements of the spherules}

\begin{figure*}
\includegraphics[width=15cm]{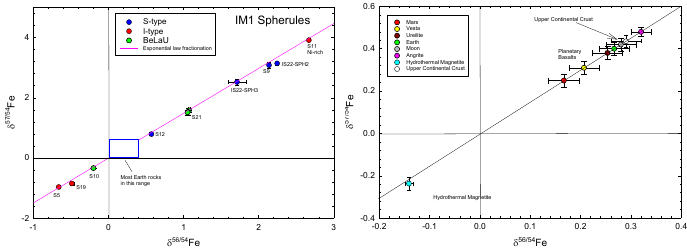}
\caption{$\delta^{57/54}$Fe versus $\delta^{56/54}$Fe for spherules in Table~\ref{tab:iron_abundances}. All spherules lie outside the box in the diagram of the left panel. This box corresponds to most terrestrial igneous rocks, with detail shown on the right panel, including planetary basalts summarized by~\citet{sossi2016iron} and~\citet{NiChabot2020}. The upper continental crust estimate is from~\citet{gong2017average}.}
\label{fig:fe_a_and_b}
\end{figure*}

The iron isotope data for the selected spherules are reported in Table~\ref{tab:iron_abundances} as $\delta^{56/54}$Fe and $\delta^{57/54}$Fe values relative to our lab standard. We also report fractionation corrected $^{57}$Fe/$^{54}$Fe ratios as $\varepsilon^{57/54}$Fe values. We measured 9 spherules for iron isotopes, 7 of them from close to IM1's path and 2 from a distant region (run 22). The $\delta^{56/54}$Fe and $\delta^{57/54}$Fe results are shown in Figure~\ref{fig:fe_a_and_b}. Note that the $\delta$-scales are relative to our Fe isotope lab standard and not IRMM-14, which has $\delta^{56/54}$Fe = 0.207 relative to our lab standard. Figure~\ref{fig:fe_a_and_b} also shows the calculated trend for mass-dependent fractionation following the exponential fractionation law of~\citet{russell1978isotope}. We also measured a hydrothermal terrestrial magnetite (OR-13) from a bismuth ore deposit associated with the Drammen Granite in Viken, Norway. The OR-13 magnetite has a $\delta^{56/54}$Fe value of -0.141 and was used together with IRMM-14 to test and monitor the methods for Fe isotope measurements. All measurements fall along the mass-dependent curve.

\begin{table}[ht]
    \centering
    \begin{tabular}{c c c c c c c c c c c}
        \toprule
        {Spherule type} & {Vial top label} & {Spherule name} & $\delta^{56/54}$Fe & $\pm1\sigma$ & $\delta^{57/54}$Fe & $\pm1\sigma$ & $\epsilon^{57/54}$Fe & $\pm1\sigma$ \\
        \midrule
        \multicolumn{9}{l}{\bf BeLaU spherules}\\
        BeLaU & S10 & IS4B-SPH1 & -0.2 & 0.002 & -0.335 & 0.011 & -0.18 & 0.14 \\
        BeLaU & S21 & IS14-SPH1 & 1.073 & 0.010 & 1.581 & 0.061 & -0.25 & 0.44 \\
        BeLaU & S21-repeat & IS14-SPH1 & 1.054 & 0.004 & 1.516 & 0.083 & -0.4 & 0.89 \\
        \multicolumn{9}{l}{\bf I-type spherules (Ni-rich)}\\
        I-type & S11 & IS4C-SPH2 & 2.668 & 0.012 & 3.911 & 0.042 & -0.6 & 0.51 \\
        \multicolumn{9}{l}{\bf I-type spherules (Ni-poor)}\\
        I-type & S5 & IS8-SPHR & -0.663 & 0.016 & -0.962 & 0.035 & 0.04 & 0.54 \\
        I-type & S19 & IS13B-SPH2 & -0.488 & 0.029 & -0.847 & 0.053 & -1.2 & 0.24 \\
        \multicolumn{9}{l}{\bf S-type spherules}\\
        S-type & S9 & IS4A-SPHS & 2.138 & 0.014 & 3.083 & 0.103 & -0.99 & 0.92 \\
        S-type & S12 & IS4D-SPH3 & 0.571 & 0.005 & 0.794 & 0.050 & -0.55 & 0.45 \\
        S-type & No label & IS22-SPH2 & 2.246 & 0.004 & 3.142 & 0.045 & -2.00 & 0.45 \\
        S-type & No label & IS22-SPH3 & 1.712 & 0.115 & 2.524 & 0.097 & -0.23 & 1.32 \\
        \multicolumn{9}{l}{\bf IRMM-14 standard and terrestrial magnetite sample}\\
        IRMM-14 & No label& standard & 0.207 & 0.009 & 0.266 & 0.044 & -0.38 & 0.63 \\
        OR-13 & No label & magnetite & -0.141 & 0.008 & -0.237 & 0.03 & -0.16 & 0.27 \\
        \bottomrule
    \end{tabular}
    \caption{Iron isotope measurements for different spherule types.}
    \label{tab:iron_abundances}
\end{table}

The mass-dependent Fe isotope effects of the spherules are well resolved and generally much larger than for typical Solar System planetary reservoir values. They are outside the range of typical values for planetary basalts from Mars, Earth, Moon, Vesta, the parent bodies of Angrites and Ureilites~\citep{sossi2016iron,NiChabot2020} and the upper continental crust estimate is from ~\citep{gong2017average}. The only solar system materials with clearly larger fractionations are the I-type cosmic spherules~\citep{engrand2005isotopic}.~\citet{engrand2005isotopic} found large, correlated, mass-dependent enrichments in the heavier isotopes of O, Cr, Fe, and Ni in I-type cosmic spherules collected from the deep sea, with $\delta^{56/54}$Fe in the range +20 to +36. They concluded that the isotopic fractionation and the I-type spherules are consistent with a Rayleigh distillation of the molten objects as they evaporate during their passage through the Earth’s atmosphere. Since the range in $\delta^{56/54}$Fe for S-type spherules were smaller (+1.4 to 3.2),~\citet{engrand2005isotopic} concluded that mass losses due to evaporation were probably small for most S-type spherules. The range of observed iron isotope fractionation in the IM1 "BeLaU"-type spherules ($\delta^{56/54}$Fe in the range -0.7 to + 2.7) is smaller than for typical I-type spherules in the~\citet{engrand2005isotopic} study and similar to their two S-type spherules. This suggests evaporation from IM1 during an airburst in the lower atmosphere of the Earth, which would suppress the magnitude Rayleigh isotope distillation effects in these spherules.

The measured $\varepsilon^{57/54}$Fe values are mostly within error of the terrestrial value. Small effects could be present, but this needs to be further investigated with more precise measurements and including $\varepsilon^{58/54}$Fe measurements. There are no well resolved nucleosynthetic variations ($\varepsilon^{57/54}$Fe values) at the level of precision obtained.


\conclusions  

The magnetic sled survey around IM1's path during the period of 14-28 June, 2023 discovered about 700 spherules of diameter 0.05-1.3 millimeters through 26 runs surveying \(0.26~\text{km}^2\). The spatial distribution of these spherules is significantly concentrated along the bolide path, as shown in Figure~\ref{fig:heatmap}. 

Mass spectrometric measurements of spherules along IM1's path shows high enrichment of Be, La and U, with extremely strong enrichment of refractory lithophile elements, very low refractory siderophile elements such as Re. Volatile elements, such as Mn, Zn and Pb, were most likely lost by evaporation during IM1's passage through the Earth's lower atmosphere. In addition, relatively large mass-dependent iron isotope variations ($^{56}$Fe/$^{54}$Fe correlated with $^{57}$Fe/$^{54}$Fe), supports evaporative losses from IM1's spherules during atmospheric passage.

Spherules with the ``BeLaU" abundance pattern shown in Figs.~\ref{fig:slide13}-\ref{fig:slide17} were found only along IM1's path and not in control regions. The ``BeLaU" elemental abundance pattern does not match terrestrial alloys, fallout from nuclear explosions~\citep{Wannier19}, magma ocean abundances of Earth, its Moon or Mars or other natural meteorites in the solar system. This supports the interstellar origin of IM1 independently of the measurement of its high speed, as reported in the CNEOS catalog~\citep{SL22a} and confirmed by the US Space Command~\footnote{\href{https://lweb.cfa.harvard.edu/~loeb/DoD.pdf}{https://lweb.cfa.harvard.edu/~loeb/DoD.pdf}}.

Since IM1's spherules melted off the surface of the object, the enhanced Be abundance might represent a flag of cosmic-ray spallation on IM1's surface along a extended interstellar journey through the Milky-Way galaxy~\citep{Johnson2019,Yokoyama16,gyngard2009interstellar, hedman2019using}. This constitutes a fourth indicator of an interstellar origin to IM1, in addition to its high speed, its heavy element composition and its iron isotope ratios. 
The enhanced abundances of heavy elements may explain the high material strength inferred for IM1 based on the high ram-pressure it was able to sustain before disintegrating~\citep{SL22b}.
The high material strength inferred for IM1 can potentially be tested experimentally by assembling a material mix based on the ``BeLaU" composition~\citep{Sunder}, with proper compensation for lost volatile elements.

The ``BeLaU" abundance pattern suggests that IM1 may have originated from a highly differentiated crust of a planet with an iron core outside the solar system. In that case, IM1's high speed of $\sim 60~{\rm km~s^{-1}}$ in the Local Standard of Rest of the Milky-Way galaxy~\citep{SL22a} and the extremely large number of similar objects per star, $\sim 10^{23\pm1}$, inferred statistically for the population of natural interstellar objects it represents (cf. Fig. 3 in~\citet{SL22b}), are challenging to explain by common dynamical processes.
The ``BeLaU" overabundance of heavy elements could perhaps have instead originated from $r$-process enrichment and fragmentation of ejecta from core-collapse supernovae or neutron star mergers~\citep{Radice2018,Johnson2019,Siegel22,Fuji2023}. However, the ``BeLaU" pattern also displays $s$-process enrichment which must have a separate origin, such 
as Asymptotic Giant Branch (AGB) stars~\citep{Busso99,Bist11,Karakas12}. Another possibility is that this unfamiliar abundance pattern may reflect an extraterrestrial technological origin. These interpretations will be considered critically along with additional results from spherule analysis in future publications. 


\authorcontribution{A. Loeb served as the chief scientist of the expedition, which was coordinated by R. McCallum and funded by C. Hoskinson. Other co-authors contributed to various aspects of the expedition, the collection of materials with the magnetic sled  and the analysis of these materials.} 

\competinginterests{No competing interests.} 


\begin{acknowledgements}
We thank C. Hoskinson for funding the expedition and the Galileo Project at Harvard University for administrative and research support. We are also grateful to Morgan MacLeod, Hamsa Padmanabhan and John Raymond for helpful comments on the manuscript. 
\end{acknowledgements}

\bibliographystyle{copernicus}
\bibliography{refs.bib}

\clearpage
\appendix

\section{Elemental data normalized to CI chondrites from the Harvard Laboratory.} 
\label{sec:elemental_data}

\begin{figure}[h]
  \includegraphics*[width=\textwidth, viewport=0 449 250 700, clip]{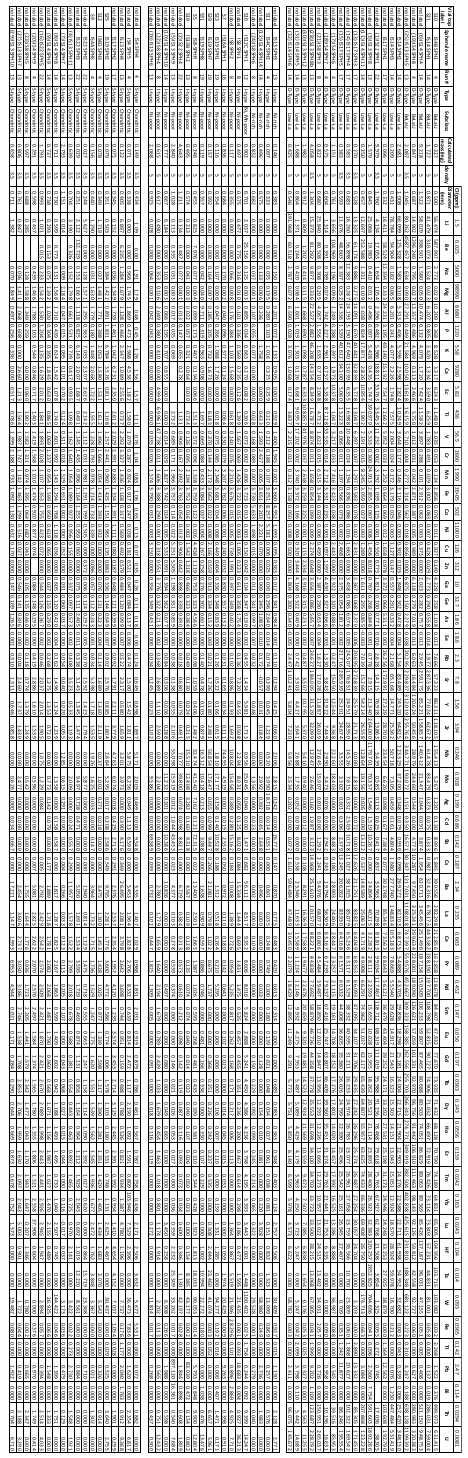}
  \caption{Elemental data normalized to CI chondrites. The CI normalizing values are given on top of the table. (Part 1 / 3)}
  \label{fig:IM1_Tables_S1_S2_part1}
\end{figure}

\begin{figure*}
  \includegraphics*[width=\textwidth, viewport=0 196 250 450, clip]{figs/appendix/IM1_Tables_S1_S2_rotated.pdf}
  \caption{Elemental data normalized to CI. The CI normalizing values are given on top of the table. (Part 2 / 3)}
  \label{fig:IM1_Tables_S1_S2_part2}
\end{figure*}

\begin{figure*}
  \includegraphics*[width=\textwidth, viewport=0 0 250 197, clip]{figs/appendix/IM1_Tables_S1_S2_rotated.pdf}
  \caption{Elemental data normalized to CI. The CI normalizing values are given on top of the table. (Part 3 / 3)}
  \label{fig:IM1_Tables_S1_S2_part3}
\end{figure*}

\section{Interstellar process notes: spherule recovery on board M/V Silver Star}    
\label{app_a}
\smallskip

\subsection{ON DECK}

\subsubsection{Collection of components}

\begin{itemize}
\item 1x Shop vacuum
\item 1x Magnetic sled
\item 1x 2 L fragment collection containers
\item 1x Hose with adjustable spray nozzle and water source
\item 1x Logbook and pencil
\item 5x Tweezers and fragment recovery implements
\item 30 L Collection container, labeled
\end{itemize}

\subsubsection{Process notes}
\begin{itemize}
\item Remove all large fragments by hand using tweezers and recovery implements into 2 L fragment collection container.
\item Shop vacuum all remaining sediment from sled into shop vacuum container, adding 10-50 mL of water per magnetic collection disk to collection to provide additional conveyance medium for sediment and fragments. Take care to not directly spray the magnetic sled, or spray only at a highly oblique angle, to ensure sled sediment and fragments are not lost.
\item Empty all captured sediment and fragments from shop vacuum container into 30 L collection container.
\item Rinse shop vacuum container into 30 L collection container.
\item Rinse shop vacuum hose into 30 L collection container.
\item Collection container sent to Lab setup Station 1.
\item Note time of recovery and sled run number in logbook.
\end{itemize}

\subsection{BRIDGE LAB:  Wet magnetic separation, Station 1}

\subsubsection{Components}
\begin{itemize}
\item 1x roll of tape
\item 4x quartered pieces of paper
\item 1x 30 L Collection container filled with sediment and fragments
\item 1x Magnetic plunger for wet use
\item 1x 250 mL measuring cup
\item 1x "Magnetic particle sieve" sieve (149 micron screen), labeled GP-1100
\item 1x 30 L "Magnetic tailings" container, labeled
\item 1x Paintbrush
\item 50 mL fresh water in measuring cup
\end{itemize}

\subsubsection{Process notes}
\begin{itemize}
\item Collection container placed at Station 1.
\item Tape a piece of paper in front of the Collection bucket and name the Collection bucket with the following convention: "[Run Number] - Collection".
\item Add 50 mL of fresh water to 250 mL measuring cup.
\item Agitate sediment resting at the bottom of the collection container using the paintbrush, making a homogenous aqueous slurry.
\item Hold magnetic plunger for wet use into 30L collection container, attracting magnetic sediment.
\item Remove magnetic plunger and place in 250 mL measuring cup.
\item Release magnetic plunger when submerged in 50 mL of fresh water, releasing magnetic particles into measuring cup.
\item Agitate the 50 mL of water and magnetic particles with brush, creating homogenous aqueous slurry.
\item Water from 250 mL measuring cup strained through magnetic particle sieve into 30 L magnetic tailings container.
\item Tape a piece of paper in front of the magnetic particle sieve with the following convention: "[Run Number] - MAG";
\item Keep all water in magnetic tailings container (contains sub-150 micron magnetic tailings).
\end{itemize}

\subsection{BRIDGE LAB: Wet non-magnetic particle separation, Station 2}

\subsubsection{Components}

\begin{itemize}
\item 1x 30 L Collection container with sediment and fragments, less magnetic particles.
\item 250 mL measuring cup
\item 1x Paintbrush
\item 1x "Non-magnetic particle sieve" sieve (300 micron screen)
\item 1x 30 L "Non-magnetic tailings" container
\item 1x Non-magnetic tailings log and pencil
\end{itemize}

\subsubsection{Process Notes}
\begin{itemize}
\item Collection container placed at Station 2.
\item Agitate sediment resting at the bottom of the Collection container using the paintbrush, making a homogenous aqueous slurry.
\item Submerge 250 mL measuring cup into Collection container, filling the measuring cup with the aqueous slurry.
\item Pour the slurry through the Non-magnetic particle sieve into the Non-magnetic tailings container.
\item Tape a piece of paper in front of the Non-magnetic particle sieve and name it with the following convention: "[Run Number] - NMAG".
\item Keep all water in Non-magnetic tailings container (contains sub-150-micron magnetic tailings).
\item Every 3 hours, remove 750 mL of fresh water off of the top of the Non-magnetic tailings container. (The sediment will settle to the bottom, and you will need room for additional slurry to be added to the Non-magnetic tailings container.)
\item Indicate the time in the Non-magnetic tailings when the removal process takes place.
\end{itemize}

\subsection{BRIDGE LAB: Sieve drying, Station 3}

\subsubsection{Components}

\begin{itemize}
\item 1x Non-magnetic particle sieve; sieve (300 micron screen) with sediment
\item 1x Magnetic particle sieve; sieve (150 micron screen) with sediment
\item 2x small trays that allow for airflow under the sieve
\item 10x paper delicate task wipers on underside to assist drying.
\end{itemize}

\subsubsection{Process notes}
\begin{itemize}
\item Place the "Non-magnetic particle sieve" and the "Magnetic particle sieve" on the small trays that allow for airflow both above and over the sieves.
\item Let rest for 5-8 hours, periodically blotting the underside of the sieves with delicate task wipers, removing excess moisture.
\end{itemize}

\subsection{BRIDGE LAB: Oven firing, Station 4 (Optional, when pressed for time by incoming sled)}

\subsubsection{Components}

\begin{itemize}
\item 1x oven
\item 1x metal cupcake tray
\item 1x hot pad
\item 1x oven mitt
\item 1x timer
\item 1x tweezers or scraping implement
\item 2x paper tags indicating sample names
\item 2x paper cup for cupcake tray
\item Magnetic sediment
\item Non-magnetic sediment
\end{itemize}

\subsubsection{Process notes}

\begin{itemize}
\item When sufficiently dry so that the sediment no longer adheres to the sieve, remove from the sieve and place in paper cup for cupcake tray.
\item Place one paper cup filled with non-magnetic sediment and one paper cup filled with magnetic sediment in the metal cupcake tray.
\item Place metal cupcake tray in the oven set to 135 degrees Celsius, for 30 minutes. Set timer for 30 minutes.
\item Move paper tags for each sample to station 4 indicating each sample is now drying in the oven.
\item Remove tray from oven using oven mitt after oven firing time has elapsed.
\item Transfer tray to Station 4, place on hot pad.
\item Optional: using thermal camera, identify hot spots in the sediment which will indicate metallic substances with a higher specific heat. These hot spots are more likely to contain spherules.
\end{itemize}

\subsection{BRIDGE LAB: Dry magnetic separation, Station 5}

\subsubsection{Components}

\begin{itemize}
\item 1x metal tray with raised sides
\item 1x small paintbrush, 3mm width
\item 1x large paintbrush, 15 mm
\item 1x magnetic plunger for dry use
\item 1x tweezers with fine point
\item 2x paper tags indicating sample names
\item 3x sheets of white paper
\item Magnetic sediment
\item Non-magnetic sediment
\end{itemize}

\subsubsection{Process notes}

\begin{itemize}
\item Remove sediment either from Station 3, sieve drying, or Station 4's cupcake tray and place on separate pieces of white paper, sheets 1 and 2. Do not mix these two samples.
\item Move the paper tags indicating the name of each sieve over to Station 5, taking care that each sample's tag is moved to Station 5 correctly.
\item Using a marker, label the samples directly on the white sheets of paper 1 and 2 with "[Run Number] Magnetic" and "[Run Number] Non-magnetic". Align paper tags with the sheets of paper.
\item Using the tweezers with a fine point, spread out the magnetic sediment on sheet 1. When dry, the sediment should not clump together.
\item Using the tweezers with a fine point, spread out the non-magnetic sediment on sheet 2. When dry, the sediment should not clump together.
\item Using magnetic plunger for dry use, hover over magnetic sediment on sheet 1, and transfer all magnetic particles to sheet 3. Agitate the magnetic sediment on sheet 1 to change its distribution. Repeat this process until all magnetic particles have been transferred from sheet 1 to sheet 3.
\item When all magnetic particles have been transferred to sheet 3, lift sheet 3 and bend the paper to aggregate all magnetic sediment in the center of the sheet.
\item Transfer all remaining material on sheet 1 to sheet 2, combining this material with sheet's non-magnetic sediment. (The remnants on sheet 1 are non-magnetic material.)
\end{itemize}

\subsection{BRIDGE LAB: Bottling and labeling, Station 6}

\subsubsection{Components}

\begin{itemize}
\item 1x sheet of paper with Magnetic sediment
\item 1x sheet of paper with Non-magnetic sediment
\item 1x label maker
\item 1x tray for newly created sample vials
\item 2x transparent glass vials
\item 2x paper tags indicating sample names
\end{itemize}

\subsubsection{Process notes}

\begin{itemize}
\item Fold the sheet of paper with magnetic sediment in half, length-wise.
\item Lift the sheet of paper with magnetic sediment and pour the material into a transparent glass vial, using the fold of the paper as a channel to convey the material into the vial.
\item Lift the sheet of paper with non-magnetic sediment and pour the material into a transparent glass vial, using the fold of the paper as a channel to convey the material into the vial.
\item Create a label with the label maker for the vial now filled with Magnetic sediment, using the naming convention "[Run Number] - MAG" and place the label on the vial.
\item Create a label with the label maker for the vial now filled with Non-magnetic sediment, using the naming convention "[Run Number] - NMAG" and place the label on the vial.
\item Place both vials in the tray for samples that have yet to be searched (brand new, recently created).
\item Move each sample's paper tag from Station 5 to Station 6 to clearly indicate what samples are now ready for examination in the next stage.
\end{itemize}

\subsection{BRIDGE LAB: Optical search, Station 7}

\subsubsection{Components}

\begin{itemize}
\item 1x sample
\item 1x Takmly Digital Microscope USB HD Inspection Camera 50x-1000x displays to laptop
\item 2x Elikliv EDM9 7'' LCD Digital Microscope 1200X with photographic capacity
\item 1x tray for (7A1) samples that have yet to be searched (brand new, recently sorted)
\item 1x tray for (7A2) samples that have yet to be searched
\item 1x tray for (7B) samples that have been searched
\item 1x tray for (7C) new vials containing spherules
\item 1x spherule logbook and pencil
\end{itemize}

\subsubsection{Process Notes}

\begin{itemize}
\item Take sample from (7A1) collection of samples that have yet to be searched (brand new, recently sorted) OR (7A2) samples that have yet to be searched to stations with microscope.
\item Examine sample under the microscope to look for small spherical objects.
\item Remove any spherical, metallic objects above 100 microns and place in separate container.
\item Study spherule under another microscope, separately.
\item Record diameter of spherule in microns and photograph.
\item Place spherule in separate container, label with the following naming convention: "[Run Number - Spherule]".
\item Place new vial in tray for (7C) new vials containing spherules.
\end{itemize}

\section{Preliminary laboratory analysis at UC Berkeley} 
\label{app_b}

\begin{figure*}[t]
\includegraphics[width=12cm]{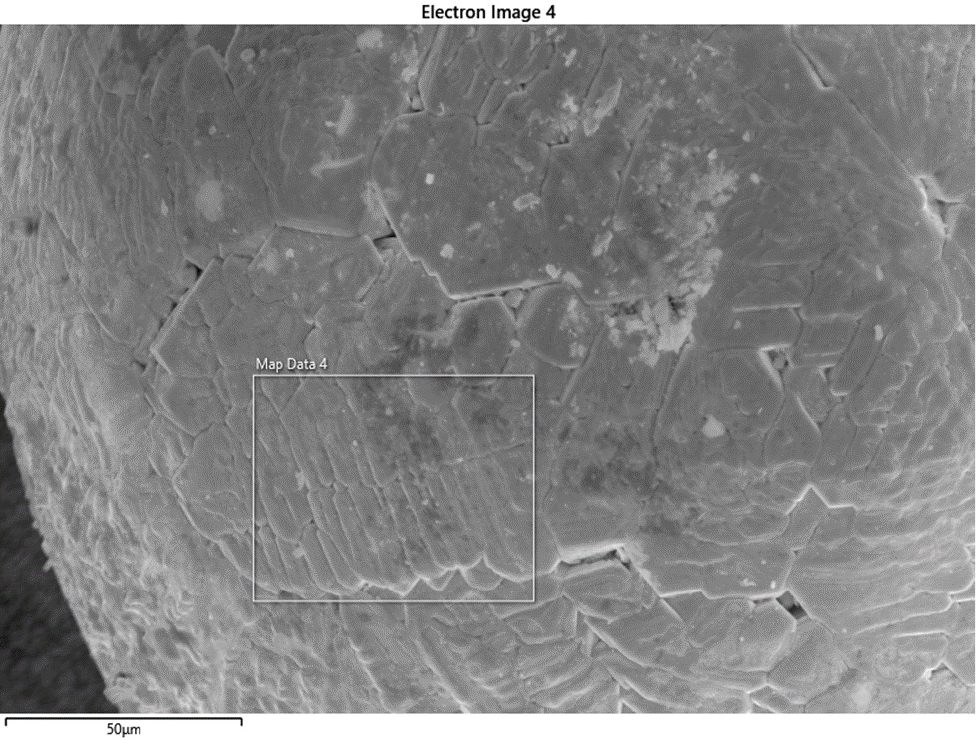}
\caption{Spherule 14 from run 8, showing full-grown dendritic structure. Different regions show various sizes of grain structure, probably due to differences in the thermal history (solidification rate, grain nucleation).}
\label{fig:ucb_fig12}
\end{figure*}

Scanning Electron Microscope and Energy Dispersive X-Ray Spectroscopy (SEM-EDS) measurements were conducted at UC Berkeley on an initial inventory of spherule samples. A Scios™ 2 DualBeam™ FIB-SEM system was used to provide 3D characterization for the range of samples, including magnetic and non-conductive materials. These samples are dominated by magnetite exteriors with some external and internal dendritic and plate structures, along with various phases and morphology of wustite intergrowths, and complex oxide-alloy interfaces (see Figure~\ref{fig:ucb_fig12}).

\begin{figure*}[t]
\includegraphics[width=12cm]{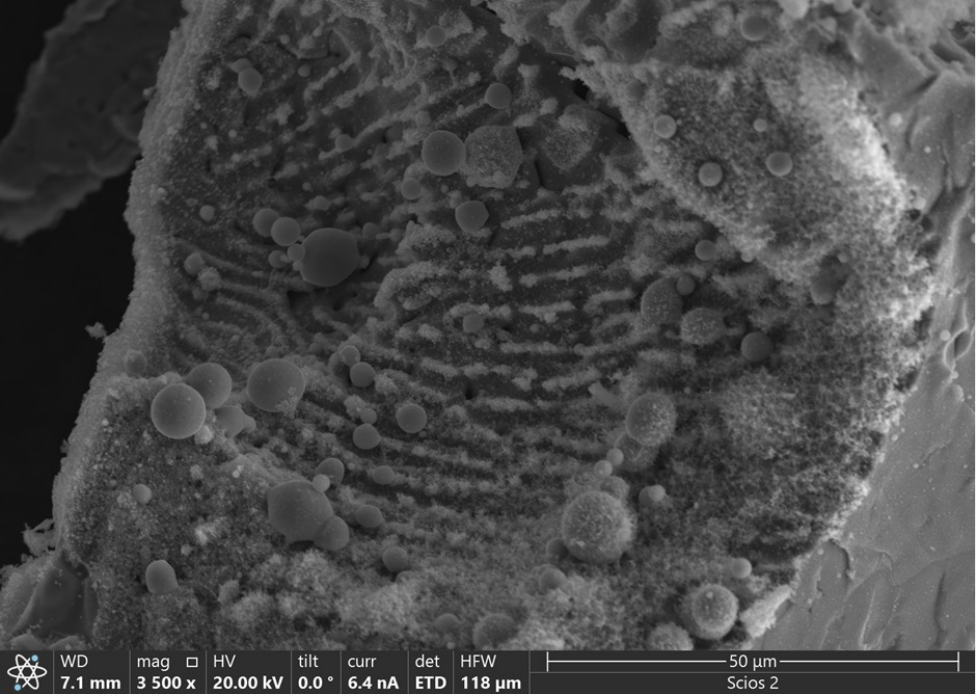}
\caption{SEM image of Spherule S4 (run 8) showing interior micro-spherules of approximately 5-10 microns in diameter.}
\label{fig:ucb_fig13}
\end{figure*}

A single sample collected far away (tens of km) from the IM1 target site contained the largest Ni
fraction (0.6\% wt), although still much lower than median Ni content for solar system
spherules~\citep{Genge17,genge2008classification,Herzog99}. For the spherules found inside IM'1 search area, the Fe-Mg-Si
ratios are consistent with other deep sea and Antarctic spherules~\citep{Rochette2008}. Additionally,
the SEM/EDS study indicates that many spherules consist of agglomerations of many smaller
spherules inside a melted matrix, a behavior consistent with previous studies~\citep{Rochette2008}. 

In general, dendritic features indicate rapid and heating processes that are consistent with atmospheric entry events. A single sample collected far away (>5km) from the IM1 target site contained the largest Ni fraction (0.6\% by weight), which is within the Ni content range for solar system spherules.  

The SEM-EDS study indicates that spherules consist of agglomerations of many smaller spherules inside a melted matrix. 

\begin{figure*}[t]
\includegraphics[width=12cm]{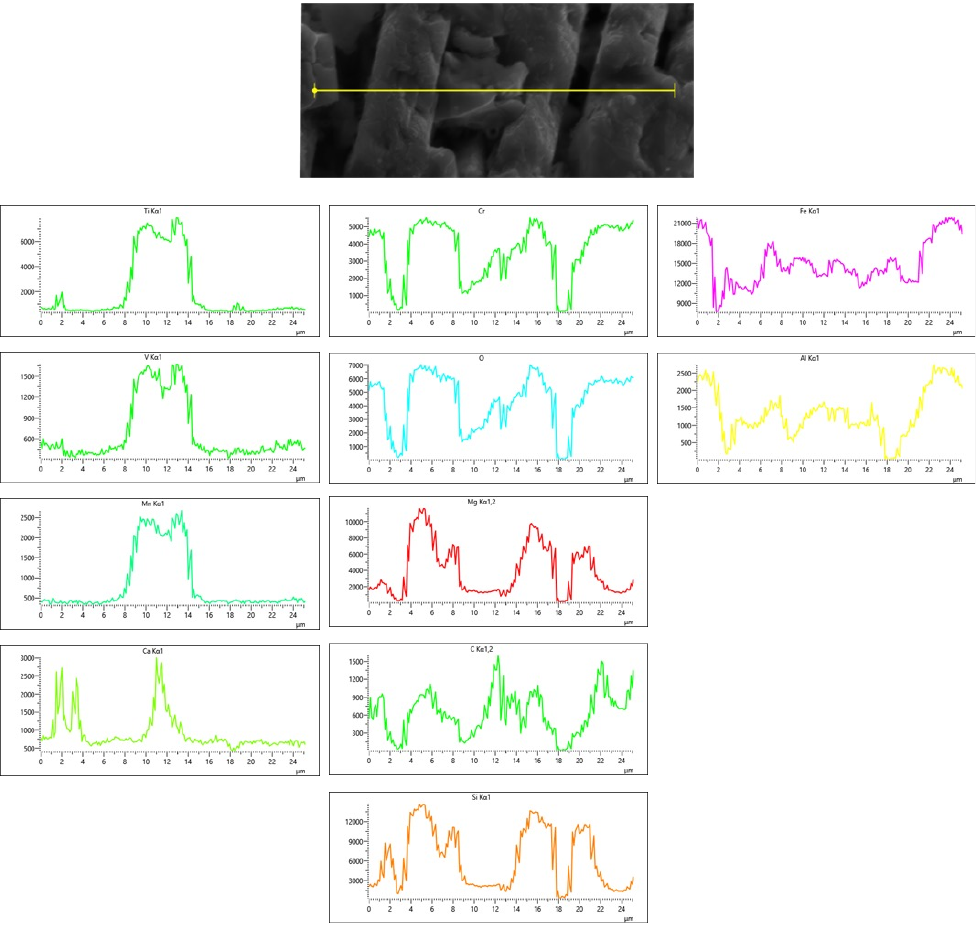}
\caption{Spherule 14 (run 8), Showing a line-scan EDS analysis across dendritic structures and precipitate composed primarily of Ti with traces of V and Mn.}
\label{fig:ucb_fig15}
\end{figure*}

Several spherules were found within a much larger (mm-scale) sized irregular
shaped Fe-rich matrix, which could indicate a possible precursor terrestrial
hydrothermal source, where previous studies have collected magnetite spherules with
interlocking plate structures~\citep{Agarwal22}. 

\begin{figure*}
\includegraphics[width=12cm]{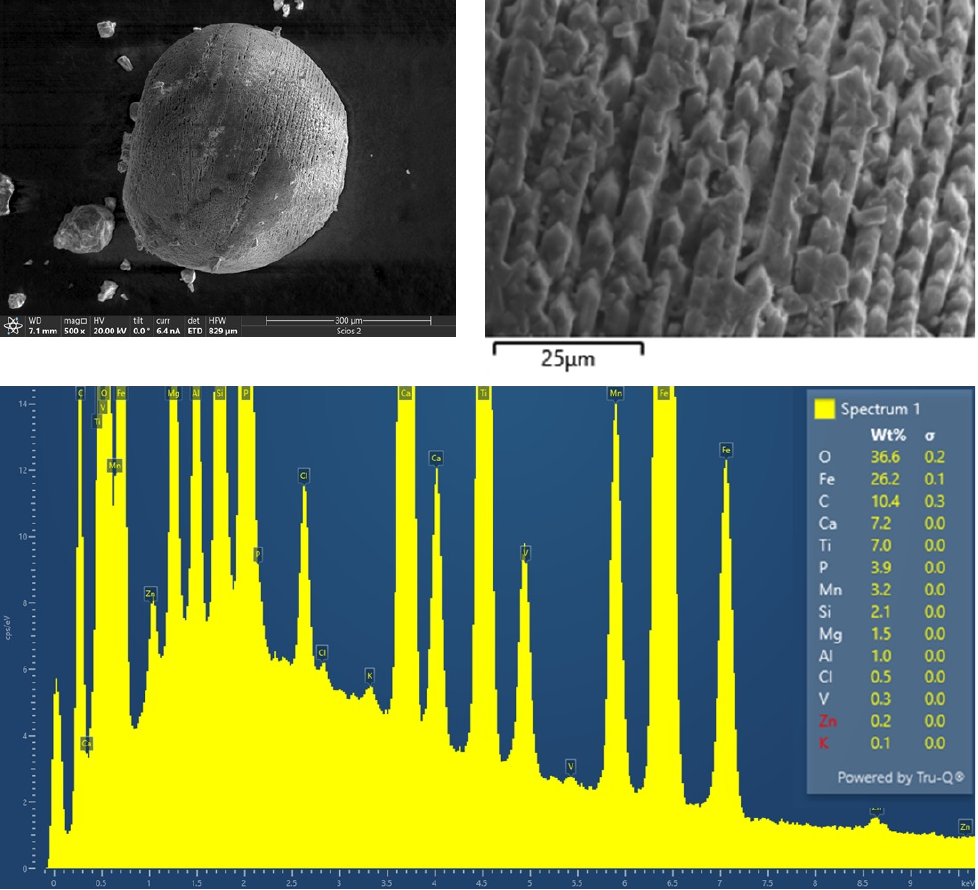}
\caption{SEM image of Spherule S4 (run 8) showing interior micro-spherules of approximately 5-10 microns in diameter.}
\label{fig:ucb_fig14}
\end{figure*}

A single cross section of a target area spherule
did not reveal any dendritic interior structure. On another target area of a spherule 
(see Figure~\ref{fig:ucb_fig14}) a line of dendritic ridges indicates Ti-rich and Mn-rich precipitates. 

\begin{figure*}
\includegraphics[width=12cm]{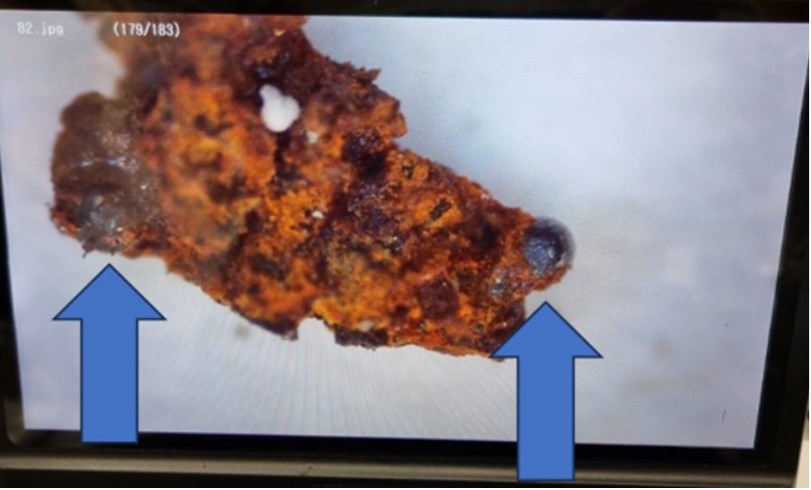}
\caption{Optical microscope image of a sample collected on run 25, showing an aggregate iron-oxide particle with embedded spherules (blue arrows). The aggregate is approximately 2mm in length}
\label{fig:ucb_fig16}
\end{figure*}

\begin{figure*}
\includegraphics[width=12cm]{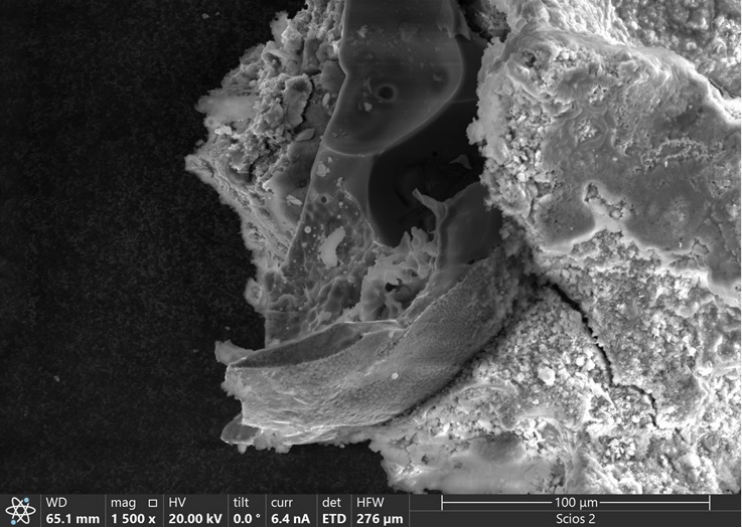}
\caption{The internal structure of spherule from run 25, embedded in Fe-rich matrix.}
\label{fig:ucb_fig17}
\end{figure*}

\begin{figure*}
\includegraphics[width=12cm]{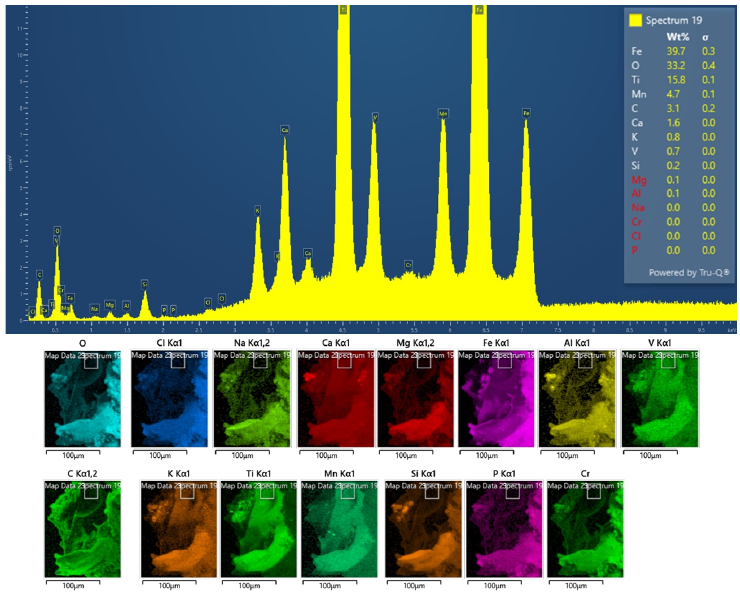}
\caption{EDS map of fractured spherule from the aggregate matrix acquired on run 25}
\label{fig:ucb_fig18}
\end{figure*}

During sample processing for SEM analysis, the spherule on the right side of Figure~\ref{fig:ucb_fig16} fractured, allowing for an internal morphology (Figure~\ref{fig:ucb_fig17}) and composition analysis. The EDS mapping of the internal structure revealed extremely high Ti content (>15\% by weight), as shown in Figure~\ref{fig:ucb_fig18}.

\begin{figure*}
\includegraphics[width=10cm]{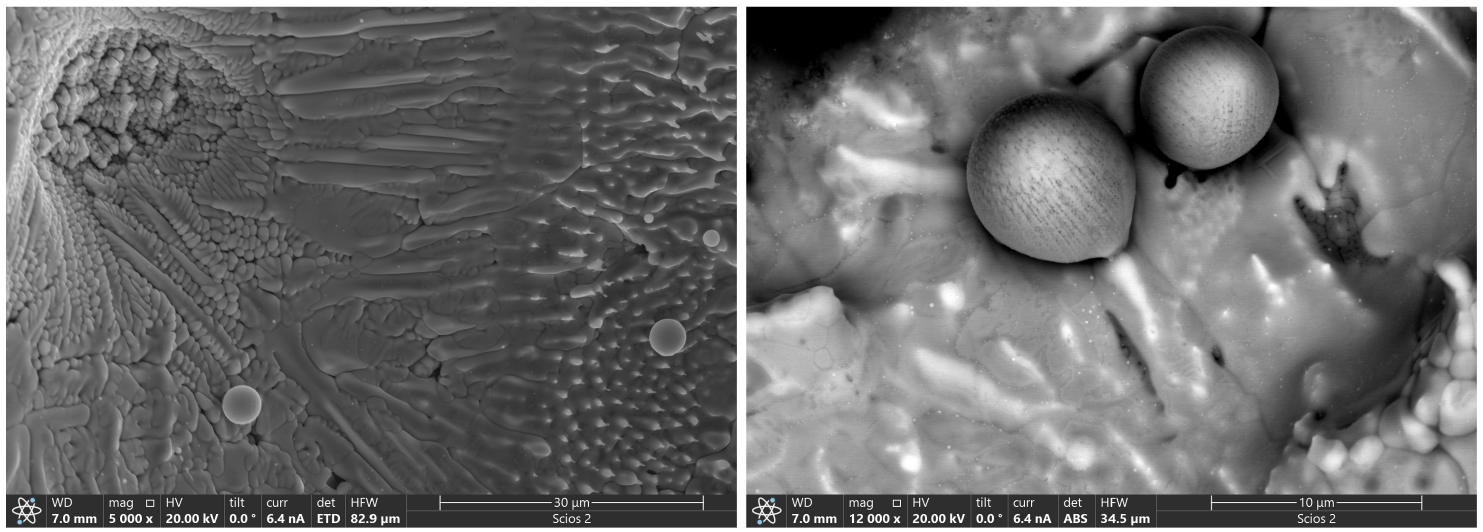}
\caption{Spherule 4 from run 8 along IM1's path, showing dendritic structure. Note several small spherical particles with fine structure are embedded in the core of larger particles.}
\label{fig:ucb_slide9}
\end{figure*}

\begin{figure*}
\includegraphics[width=10cm]{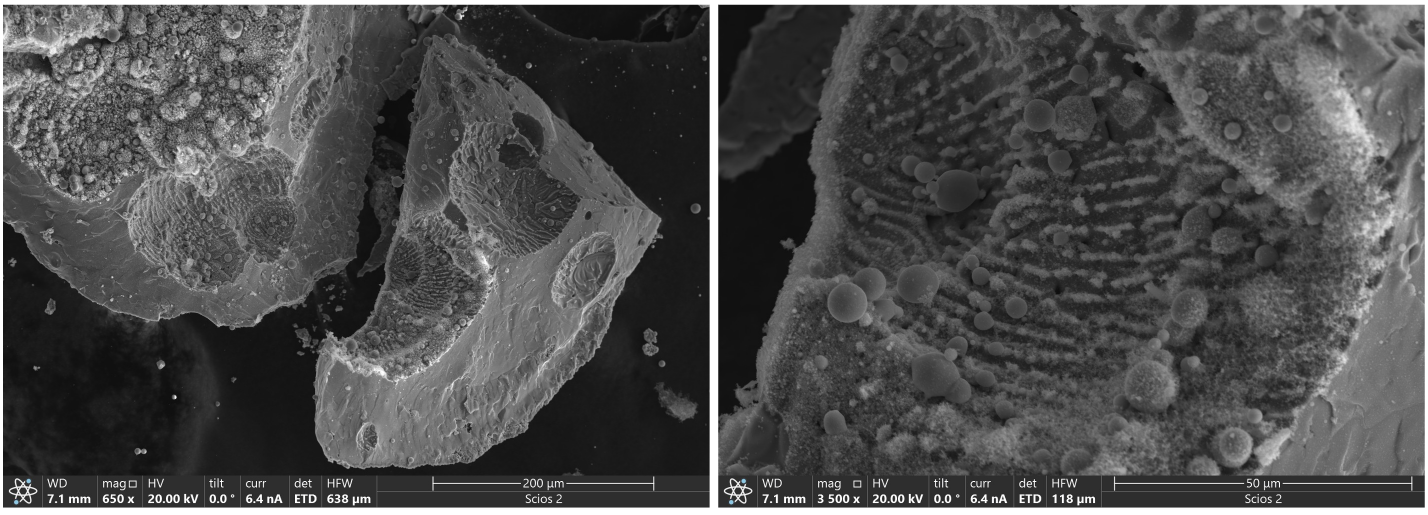}
\caption{Spherule 4 from run 8, showing many smaller-sized particles embedded in the core of the larger structure. }
\label{fig:ucb_slide10}
\end{figure*}


%
%

\noappendix       




\appendixfigures  

\appendixtables   













\end{document}